\documentclass{article}

\usepackage{microtype}
\usepackage{graphicx}
\usepackage{subfigure}
\usepackage{booktabs} %

\PassOptionsToPackage{hyphens}{url}
\usepackage{hyperref}

\usepackage[accepted]{icml2024}

\usepackage{amsmath}
\usepackage{amssymb}
\usepackage{mathtools}
\usepackage{amsthm}

\usepackage[capitalize,noabbrev]{cleveref}
\usepackage{todonotes}

\definecolor{black}{RGB}{0,0,0}
\newcommand*\negcircnum[1]{\tikz[baseline=(char.base)]{%
            \node[white,shape=circle,fill=black,draw,inner sep=1pt] (char) {\color{white}\sffamily #1};}}

\newcommand*\negcircnumred[1]{\tikz[baseline=(char.base)]{%
            \node[white,shape=circle,fill=red,draw,inner sep=1pt] (char) {\color{white}\sffamily #1};}}

\definecolor{newBlue}{RGB}{162, 210, 255}

\usepackage{adjustbox}

\theoremstyle{plain}

\theoremstyle{definition}

\theoremstyle{remark}

\newcommand{\intel}{Intel\textsuperscript{\textregistered}}
\newcommand{\intels}{Intel's\textsuperscript{\textregistered}}

\icmltitlerunning{Fortify Your Foundations: Practical Privacy and Security for Foundation Model Deployments In The Cloud}

\begin{document}

\twocolumn[
\icmltitle{Fortify Your Foundations: Practical Privacy and Security for Foundation Model Deployments In The Cloud}

\icmlsetsymbol{equal}{*}

\begin{icmlauthorlist}
    \icmlauthor{Marcin Chrapek}{eth}
    \icmlauthor{Anjo Vahldiek-Oberwagner}{ilb}
    \icmlauthor{Marcin Spoczynski}{il}
    \icmlauthor{Scott Constable}{il}
    \icmlauthor{Mona Vij}{il}
    \icmlauthor{Torsten Hoefler}{eth}
\end{icmlauthorlist}

\icmlaffiliation{eth}{ETH Zurich, Zurich, Switzerland}
\icmlaffiliation{il}{Intel Labs, Portland, USA}
\icmlaffiliation{ilb}{Intel Labs, Berlin, Germany}

\icmlcorrespondingauthor{Marcin Chrapek}{marcin.chrapek@inf.ethz.ch}

\icmlkeywords{Machine Learning, ICML}

\vskip 0.3in
]

\printAffiliationsAndNotice{}  %

\begin{abstract}

Foundation Models (FMs) display exceptional performance in tasks such as natural language processing and are being applied across a growing range of disciplines. Although typically trained on large public datasets, FMs are often fine-tuned or integrated into Retrieval-Augmented Generation (RAG) systems, which rely on private data. This access, along with their size and costly training, heightens the risk of intellectual property theft. Moreover, multimodal FMs may expose sensitive information. In this work, we examine the FM threat model and discuss the practicality and comprehensiveness of various approaches for securing against them, such as ML-based methods and trusted execution environments (TEEs). We demonstrate that TEEs offer an effective balance between strong security properties, usability, and performance. Specifically, we present a solution achieving less than 10\% overhead versus bare metal for the full Llama2 7B and 13B inference pipelines running inside \intel\ SGX and \intel\ TDX. We also share our configuration files and insights from our implementation. To our knowledge, our work is the first to show the practicality of TEEs for securing FMs.

\end{abstract}

\section{Introduction}

\begin{figure}[!t]
  \centering
  \includegraphics[width=\linewidth]{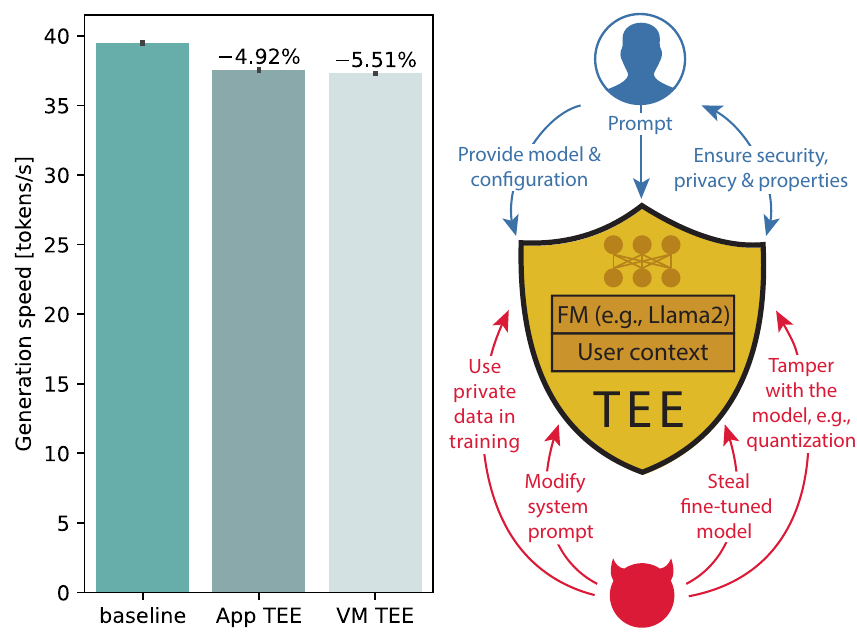}
  \caption{Examples of the types of threats our approach leveraging TEEs protects actively against. We also show our example performance results for baseline inference of Llama2 7B INT8 in two TEE implementations, a Virtual Machine (VM) and an application-based one.}
  \label{fig:child_poster}
\end{figure}

Foundation Models (FMs) are dominating the machine learning (ML) landscape. Exemplified by Large Language Models (LLMs), such as GPT-4~\cite{openaiGPT4TechnicalReport2023a}, GPT-3~\cite{brownLanguageModelsAre2020}, Llama~\cite{touvronLLaMAOpenEfficient2023}, and Llama2~\cite{touvronLlamaOpenFoundation2023}, they displayed impressive in-context learning abilities. FMs such as LLMs achieve human-like capabilities that have revolutionized many ML tasks~\cite{awaisFoundationalModelsDefining2023, zhaoSurveyLargeLanguage2023} and have been successfully applied to disciplines relying on \textit{confidential} user data such as healthcare~\cite{sallamChatGPTUtilityHealthcare2023}, finance~\cite{wuBloombergGPTLargeLanguage2023}, sentiment analysis~\cite{araciFinBERTFinancialSentiment2019}, legal cases~\cite{cuiChatLawOpenSourceLegal2023}, and document translation~\cite{kocmiLargeLanguageModels2023}. Simultaneously, the ever-increasing size of FMs has changed their deployment strategies. FMs reach billions to trillions of parameters, necessitating state-of-the-art hardware to achieve reasonable performance. However, such hardware is frequently unavailable or too expensive for small developers. This results in models not being deployed locally but offered as a service~\cite{ribeiroMLaaSMachineLearning2015} by major cloud service providers (CSPs). However, CSPs have an interest in using the data users provide for model training. Imagine a situation where you upload health-related documents after an accident for an insurance claim. The insurer uses an LLM deployed in the cloud to parse your data, put it into the correct format for a database, and check for abnormalities. In a couple of weeks, you open a publically available LLM to realize that after inputting a certain sequence of characters~\cite{carliniExtractingTrainingData2021,patilCanSensitiveInformation2023}, your name, address, social security number, prior health history, and accident details are visible. Someone stole your data, or the CSP or model owner used it for training. Such a drastic shift in data interest compared to previous ML models~\cite{xueIntellectualPropertyProtection2022} introduces new types of adversaries we discuss in Section~\ref{sec:threat_model}.

Such new adversaries create new threats. The right side of Figure~\ref{fig:child_poster} shows example attacks that FMs need to be shielded from. Protection against these threats is becoming increasingly important for FMs, which become abundant in companies and everyday life in a growing number of domains with access to multimodal~\cite{wuNExTGPTAnytoAnyMultimodal2023} user confidential data. This trend reflects a broader shift in focus toward privacy and security among governments, companies, and users~\cite{goldfarbShiftsPrivacyConcerns2012, petrescuAnalyzingAnalyticsData2018, vossEuropeanUnionData2017}. %
Apart from confidentiality, theft is another threat to FMs. The cost of obtaining the necessary datasets and engineering considerably increases the IP value of FMs. Training and fine-tuning alone can cost tens of millions of dollars~\cite{sharirCostTrainingNLP2020}. Any security breach involving FMs, including the leak of confidential data or IP, is becoming increasingly costly for CSPs (e.g., Azure, AWS, Google Cloud), model providers (e.g., Meta AI, OpenAI), and end-users (e.g., banks, hospitals). As more companies enter the space of personalized AI (e.g., Meta's AI studio or Adobe Creator), we already observe backlashes from users interested in their data~\cite{ngAdobeSaysIt}, making such threats tangible and requiring immediate attention.

The ML community approached the security and privacy issues associated with third-party evaluated DNNs~\cite{xueIntellectualPropertyProtection2022, knottCrypTenSecureMultiParty2022, dowlinCryptoNetsApplyingNeural} using mechanisms such as watermarking or user authentication. A practical alternative to these methods can be trusted execution environments (TEEs). TEEs promise strong security features at the expense of workload-dependent performance. We discuss these and other approaches in Section~\ref{sec:related_work}. We show that TEEs provide strong, measurable security properties for FMs against the threats important to industry companies deploying models in the cloud, in line with the model from Section~\ref{sec:threat_model}. Furthermore, besides providing security and privacy, TEEs can be leveraged to assure model-related properties, such as accuracy assurances or dataset content verification. We introduce in Section~\ref{sec:protections} a specific flow for providing security and privacy to FMs deployed within TEEs.

While TEEs provide strong properties, we investigate whether they matured enough to be a practical solution for ML practitioners interested in protecting their FMs. Past studies~\cite{moMachineLearningConfidential2023, akramPerformanceAnalysisScientific2021} have pointed to two issues with using TEEs: \textit{programming difficulty} and \textit{performance overheads}. We address both of these in our work. As we show in Section~\ref{sec:hardship}, we implement an FM inference pipeline within TEEs leveraging virtual machines (VMs) and library operating systems (OSs) such as Gramine~\cite{tsaiGrapheneSGXPracticalLibrary2017}. We open-source our setup for others to leverage and share the insights we learned throughout the process. To address the performance issues, we run an entire Llama2 inference pipeline within TEEs and, for the first time to the best of our knowledge, present performance numbers for such setup in Section~\ref{sec:performance}. The left part of Figure~\ref{fig:child_poster} displays our example performance results, showing that TEEs incur only 4-7\% throughput reduction as compared to up to 100s\% reported in the literature~\cite{akramPerformanceAnalysisScientific2021}. Finally, we also discuss training, GPU support, and the choice between different types of TEEs in Section~\ref{sec:discussion}.

\section{FM threat model}
\label{sec:threat_model}

\begin{figure}[]
  \centering
  \includegraphics[width=\linewidth]{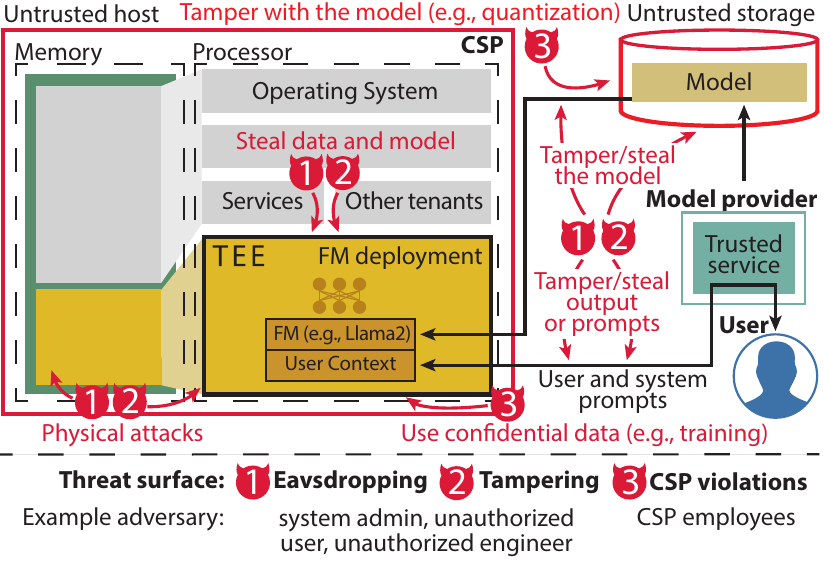}
  \caption{An overview of the threats and adversaries that exist when offloading FM deployments to the cloud, together with representative examples.}
  \label{fig:threat_model}
\end{figure}

Three main actors of interest are a part of the modern FM ecosystem: CSPs, model providers, and users. CSPs (e.g., Azure, GCP, AWS) usually only offer a service where models can be deployed and rarely work on FMs. Most FMs are developed by model providers with two forms: AI-focused (e.g., OpenAI, Anthropic) or non-AI-focused (e.g., bank, hospital, insurer). The former focuses on building and offering users proprietary general models (e.g., GPT4) as a website or API. The latter usually leverages public models (e.g., Llama) to build case-dependent proprietary models offering users specific services or interfaces to use them. This is usually achieved by fine-tuning the public models or creating Retrieval-Augmented Generation (RAG) solutions based on the company's proprietary and sensitive files, which are added to the FM query to provide additional context. From an economic perspective, it is not worth training the model from scratch for these non-AI-oriented companies. We focus our work predominantly on threats to such private company deployments on the cloud infrastructure as these have access to more confidential data than general AI-focused companies. The users in such a case can be private individuals (e.g., insured persons or patients) or company employees (e.g., internally deployed HR model). We discuss other deployment possibilities in Section~\ref{sec:other_threats}.

Examples of deployments that we focus on are a bank running an LLM parsing client statements to provide insights (e.g., how much did I spend on groceries this month), an insurer checking medical bills for abnormalities, or healthcare-provider parsing documentation (e.g., for personalized medication). Such high-value industries need IP protection for their deployed models, as not only do they constitute a competitive advantage, but they also may contain company secrets and user data. Even if the model used has not been fine-tuned, as is a vanilla public model, these industries require the confidentiality of user data. Because of these needs, the aforementioned industries cannot leverage efficient CSP scaling of advances in FMs. Promises from the CSPs that the data will not be used for other training or that the models will not be investigated are frequently insufficient guarantees.

In the above setting, we differentiate between three types of threat surfaces and their corresponding adversaries. The first two are connected with malicious actors trying to steal the model, steal user data, or disrupt the service. The third is associated with the dishonest and organized operation of the CSP. From the perspective of the model provider, the only trusted entity is the hardware itself. The end user trusts the model provider. This assumption is reasonable, as end users already do so for institutions such as banks and hospitals or their employers. The operating system, the network, the system administrators, and the CSP are not trusted. We present an overview of these threat surfaces and adversaries in Figure~\ref{fig:threat_model}. We focus only on inference and discuss training in Section~\ref{sec:training}. Finally, we do not consider a distributed setup with multiple computing hosts~\cite{ben-nunDemystifyingParallelDistributed2019}. This considerably increases the trusted computing base (TCB) and is usually unnecessary for the inference we focus on.

\textbf{\negcircnumred{1} Eavesdropping:} The first threat surface represents adversaries that try to eavesdrop on either the models or the user's confidential data. These adversaries are usually privileged in some way by either being employed by the CSP or having unauthorized access to the cluster. However, they are malicious and do not act in line with the CSP. An example of such attackers would be a rogue system administrator or another tenant within the system. They can obtain the weights or prompts by locating themselves within the network or the system. Furthermore, such adversaries could try to extract sensitive user data from the model by prompting it in the background within the cloud. Finally, the adversary might use certain input/output pairs of the model as their private training data. All these thefts offer potentially large gains with relatively low risks, as detecting such actions at the scale is nontrivial. We do not consider whole model extraction attacks~\cite{tramerStealingMachineLearning2016} as a big threat as such attacks scale poorly for large models such as FMs. %

\textbf{\negcircnumred{2} Tampering:} The second type of threat surface corresponds to adversaries not interested in stealing the models but disrupting the service through modifying critical data. Similarly to the first type, they are also malicious and do not act in line with the CSP. Again, examples include system administrators and other malicious software on the system that is trying to disrupt or hinder the program. They might tamper with the model's weights randomly to lower the prediction accuracy and change the output or the user context, which usually results in service disruption. 

\textbf{\negcircnumred{3} CSP violations:} While similar to the second type, we specified the third type of threat surface separately due to the uniqueness of the organized approach of the actors operating on it. Unlike the prior two, the third adversary is not malicious. It is the CSP that is interested in violating different contract agreements by, for example, modifying the deployed model or using the data to train other models. The goal of the modifications might be to improve the performance of their runtimes and save money (e.g., automatic quantization) or reduce the accuracy to make the users migrate to better and newer models. Such modifications without the consent of the users happen in practice~\cite{chenHowChatGPTBehavior2023}. This adversary is new and did not exist in the prior ML models as their data was rarely in a unified format, and their runtimes were considerably smaller and faster, not requiring as much optimization.

\section{Methods to protect FMs}
\label{sec:related_work}

Broadly, three approaches can be applied to protect against security threats in FMs: ML methods, cryptographic methods such as Homomorphic Encryption (HE) and multiparty computation (MPC), and Confidential Computing (CC)~\cite{mulliganConfidentialComputingBrave2021}. 

As noted in the literature~\cite{xueIntellectualPropertyProtection2022}, current ML IP protection methods lack in the space of actively protecting against model theft and instead focus predominantly on model verification and passive protections of already stolen models. The task there is to determine whether an output is coming from a given predefined model. One of the approaches is to use signatures embedded in the model to then submit multiple inputs and using outputs verify that the model is the one that was promised~\cite{laoDeepAuthDNNAuthentication2022}. This partially covers threats two and three from our threat model. Other methods include approaches such as passport~\cite{fanRethinkingDeepNeural2019} or backdoor~\cite{xueActiveDNNIP2020} based user authentication, and watermarking~\cite{szyllerDAWNDynamicAdversarial2021, boenischSystematicReviewModel2021}, in which a watermark is included in the model's output or weights, allowing for ownership verification. 

While these protect against certain attacks, the threat model for ML methods does cover the threats we show in Section~\ref{sec:threat_model}. Most importantly, they do not provide exhaustive and measurable security properties, making it risky for companies to rely purely on them, considering the cost of losing confidentiality or theft of IP. Additionally, ML methods have other crucial issues. They frequently require expensive FM retraining, change the accuracy of the model, do not secure the confidentiality of user prompts~\cite{xueIntellectualPropertyProtection2022}, and cannot be combined together~\cite{szyllerConflictingInteractionsProtection2023}. Cryptographic approaches such as HE and MPC address these issues with strong cryptographic protocols.

HE allows to conduct mathematical and logical operations on encrypted data without decrypting it~\cite{acarSurveyHomomorphicEncryption2018}. HE has been explored in the context of DNNs~\cite{dowlinCryptoNetsApplyingNeural, leePrivacyPreservingMachineLearning2022, woodHomomorphicEncryptionMachine2020}. However, with the exception of a few structured examples~\cite{chrapekHEARHomomorphicallyEncrypted2023, burkhalterTimeCryptEncryptedData2020}, the current state-of-the-art HE is not practical. HE approaches do not provide integrity protections (threats two and three).  HE operations on encrypted data can also have up to 10,000x performance and size overheads, taking minutes to conduct simple MNIST inference~\cite{dowlinCryptoNetsApplyingNeural} and making FM inference intangible. MPC is close to HE and has similar practicality issues but involves multiple computing parties~\cite{viandMarbleMakingFully2018a}.

CC offers a more practical and time-tested alternative in the form of TEEs by approaching the problem using strong security primitives implemented in hardened hardware. Compared to HE and MPC which rely on obscuring the data and functions, TEEs offer a secure and isolated environment frequently called an \textit{enclave}. Users can verify enclaves in a secure, hardware enabled process called \textit{attestation}. TEEs ensure the confidentiality and integrity of a running program and its data, and protect against external and privileged attackers such as system administrators. TEEs ensure these adversaries cannot access or modify the contents of the memory of the running programs~\cite{sabtTrustedExecutionEnvironment2015} such as the weights or user confidential data, mitigating risks on the user side and reducing responsibility on the CSP side. TEEs have been implemented by the academic community and industry~\cite{schneiderSoKHardwaresupportedTrusted2022}. Only the latter are available widely on CSP platforms with examples such as AMD's Secure Encrypted Virtualization-Secure Nested Paging (SEV-SNP)~\cite{kaplanAMDSEVSNPStrengthening}, \intels\ SGX~\cite{mckeenInnovativeInstructionsSoftware2013, hoekstraUsingInnovativeInstructions2013, costanIntelSGXExplained2016} and TDX~\cite{chengIntelTDXDemystified2023}, ARM's TrustZone~\cite{pintoDemystifyingArmTrustZone2019} and CCA~\cite{liDesignVerificationArm2022}. All of these are CPU-based, but accelerators are also entering the space with notable example of Nvidia~\cite{nertneyConfidentialComputeNVIDIA2023}. We discuss their current status in Section~\ref{sec:gpu_perf}.

As we show in Section~\ref{sec:protections}, TEEs outperform ML methods by providing real-time, strong, measurable, and active protections against adversaries without any model modifications, ensuring that the model, data, and runtime environment remain secure and tamper-proof. While HE and MPC provide confidentiality, TEEs provide more than just that. They offer integrity checks, attestation, and runtime protections, all of which are absent in HE and MPC. TEEs are more suitable for scenarios where performance (Section~\ref{sec:performance}), real-time inference, and ease of use are essential, such as in healthcare, finance, and cloud-based AI services.

TEEs have been investigated in the past for protecting ML models~\cite{moMachineLearningConfidential2023}. Yet most of these approaches offload only parts of the models to the TEE, usually providing weaker notions of security and claiming the low TEE performance (hundred times slowdown~\cite{akramPerformanceAnalysisScientific2021}) as the reason. For example, Slalom~\cite{tramerSlalomFastVerifiable2019} would offload all linear layers to the GPU with a probabilistic algorithm guaranteeing some security. Such an approach does not resolve the accessibility issue of TEEs and makes it even harder to use with ML models. Furthermore, none of these previous works explored FMs and focused on simpler models, such as VGG16 or MobileNet, as the necessary model changes are large. %

We address these shortcomings and, compared to previous approaches securing only certain model stages, show that offloading whole FMs to modern TEEs is practical. In Section~\ref{sec:protections}, we address TEE accessibility, show the exact control flow of how to implement an entire pipeline for FMs in TEEs, and discuss the provided security. In Section~\ref{sec:hardship} we describe our implementation and the tools we leverage. In Section~\ref{sec:performance}, we show the performance of our approach and the cost of the achieved security, which is similar or lower to the one noted in the literature.

\section{Establishing trust in the FM deployment}
\label{sec:protections}
We cannot rely on techniques suggested in prior work~\cite{leePrivacyPreservingMachineLearning2020} to establish trust in the deployment since the involved parties are more complex and deployment options differ. The majority of CSPs now offer TEE-based instances, sometimes referred to as confidential VMs (CVMs). While deploying a TEE on a CSP infrastructure can be done with a click of a button, the whole purpose of using a TEE is lost unless it is verified to be a true TEE. As mentioned earlier, TEEs support attestation~\cite{birkholzRemoteATtestationProcedureS2023} that allows them to prove to a remote party that they are indeed a TEE with the correct software state using a generated quote. The verifier can check whether the TEE signer is an actual TEE, and the hash of the code running in a TEE. The model providers can use the CSP attestation service or an independent third-party verifier like Intel Tiber. We discuss the security aspect of this choice in the second part of this section. 

\begin{figure*}[]
  \centering
  \includegraphics[width=\linewidth]{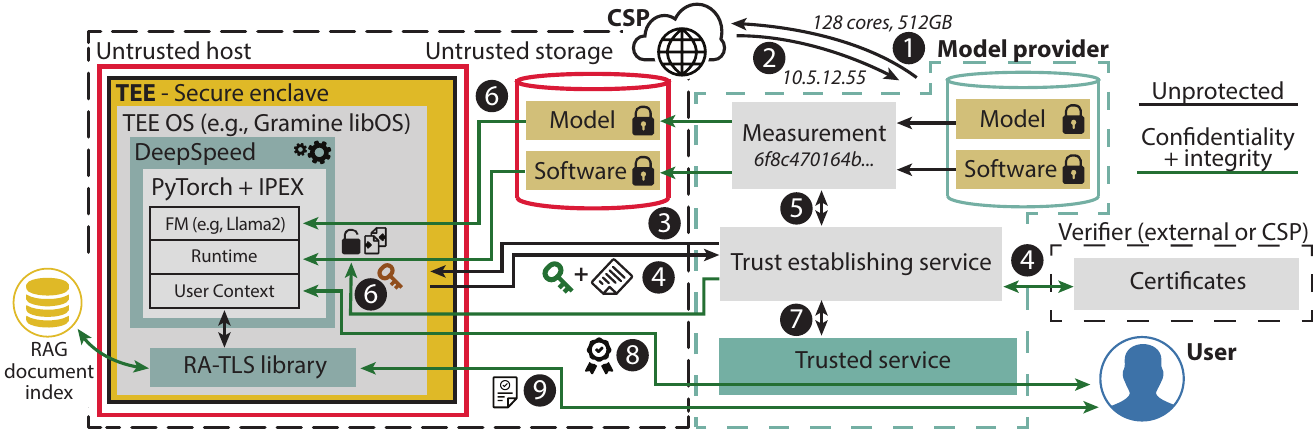}
  \caption{An overview of a flow to secure FMs relying on properties of TEEs. Here, we assume a secure enclave on an untrusted host operated by the CSP and running some kind of an OS that supports its security features. Green lines show communication channels that are protected using confidentiality and integrity in some way (e.g., TLS or encrypted storage). Black lines are unprotected.}
  \label{fig:protections}
\end{figure*}

We assume there is a model provider who wants to create a service using an FM that is fine-tuned on sensitive proprietary data or handles confidential user data that requires protection against the threats defined in Section~\ref{sec:threat_model}. Figure~\ref{fig:protections} shows the overview of the steps the model provider can follow:

\begin{itemize}
    \itemsep0em
    \item[\negcircnum{1}] The model provider requests a TEE instance using the CSPs interface specifying the software that will be run on the instance (e.g., VM image). 
    \item[\negcircnum{2}] The CSP allocates the TEE instance and provides the model provider with the associated IP needed to connect to the instance.
    \item[\negcircnum{3}] The model provider requests an attestation report from the TEE instance, which includes a public key for any subsequent communication between the model provider and the TEE.
    \item[\negcircnum{4}] The model provider leverages a verifier service to check that the running instance is a valid TEE.
    \item[\negcircnum{5}] The model provider can verify whether the TEE is running appropriate software and firmware using expected values generated before starting the TEE in the cloud.
    \item[\negcircnum{6}] The model provider provisions the encryption keys for the TEE, allowing it to load all software and the encrypted model.
    \item[\negcircnum{7}] The model provider can then expose the TEE instance running the FM as part of their service.
    \item[\negcircnum{8}] An end-user can open a direct secure channel with the TEE instance and submit inference prompts leveraging a public certificate offered by the model provider.
    \item[\negcircnum{9}] An end-user can also establish that the service is running in a TEE by verifying the attestation. This can happen transparently during secure channel opening if the users leverage a protocol such as Remote Attestation TLS (RA-TLS)~\cite{knauthIntegratingRemoteAttestation2019}. RA-TLS in such a setup requires granting user access to a verifier service.
\end{itemize}

\subsection{Protections against threats}
\label{sec:flow_threats}
The above flow protects against the attack vectors defined in Section~\ref{sec:threat_model}. We discuss how this is achieved and how TEEs offer an advantage over other methods for different threats.

\textbf{Type one and two adversaries:} Type one and two adversaries are protected against using the principal integrity and confidentiality properties of TEEs. The model (step \negcircnum{6}) and users' data (step \negcircnum{8}) are protected from access by privileged and non-privileged users of the data center. TEEs prevent unauthorized adversaries from reading plaintext data out of the TEE through, for example, garbling the output or raising memory faults. TEEs also prevent unauthorized adversaries from modifying code/data within the TEE. Some TEEs will simply crash after their data is modified in this manner, notifying the user and closing the connection. Because the user uses TLS for communication and optionally conducts a provably secure attestation, no listener on the communication ports can understand the contents of the sent messages, blocking effective tracking of the input and output.
 
\textbf{Type three adversary:} TEEs can also assert certain assurances that other methods struggle with. By conducting the attestation mentioned previously and because of TEE's strong integrity protection, the model provider can ensure they deploy the models they intended without any modifications and usage of their data. They compare the expected secure hash value of the agreed model with the secure hash value coming from the enclave. Such comparison provably ensures the CSP and its employees adhere to their part of the agreement and do not modify the models, no matter how expensive they are to process. TEEs also provide privacy of users' prompts, where the attestation protects against data or model leakage, ensuring that users' data is provably not used for training purposes. Such strong protection can be important for legally imposed standards in industries such as healthcare or finance. This allows for a unique method of ensuring the quality of service and eliminates the third type of adversary.

\textbf{Combining with other protection methods: } Businesses have varying needs and sometimes have deployments susceptible to other threats. In such cases, TEEs have another distinct advantage over ML methods. Existing ML protection mechanisms have been shown to eliminate each other's benefits when combined together~\cite{szyllerConflictingInteractionsProtection2023}. TEEs do not modify the model and, thus, are not susceptible to such an issue. TEEs can be joined with at least one other mechanism reinforcing other passive or active ML protections that might defend against a specific threat, such as model extraction. We believe TEEs comprise a secure baseline foundation for any FM deployment that does not interfere with other approaches.

\subsection{Deployment and threat considerations}
We discuss how, from a practical perspective, the above flow can be leveraged in deploying other services and how the deployment decisions influence its security.

\textbf{RAG:} The above flow also applies to retrieval-augmented generation (RAG)~\cite{lewisRetrievalAugmentedGenerationKnowledgeIntensive2020}. In RAG, FMs query a document index that responds to prompts. This directly maps to our above flow with exchanged actors. The model provider would follow the steps \negcircnum{1} to \negcircnum{7} to deploy an additional RAG service. The details of this service, such as the communication certificate, are then shared with the inference service. The inference service would open a direct, secure connection with the RAG service. Leveraging remote attestation during deployment, encrypted storage, and secure communication ensures the provided RAG documents are not maliciously changed or stolen, similarly to how we described for an inference service in Section~\ref{sec:flow_threats}.

\textbf{Load balancers and gateways:} Frequently, the user would not have direct access to an instance running on the cloud but would be located behind a load balancer or a gateway. Such mechanisms could isolate the user from the ability to attest the TEE if the TLS sessions are terminated at these points.

\textbf{Choice of verifier service:} CSPs offer attestation services for TEEs running in their deployment. To establish trust in these offerings, they should be run inside a TEE, which is attested by a trusted 3rd party. Furthermore, the attestation service implementation needs to be validated. Such offerings allow CSPs to handle attestations for TEE deployments on their infrastructure, with 3rd party required only for the few attestations of the TEEs running CSP attestation service. To further reduce the reliance on the CSP, one can also completely resort to 3rd party attestation services.

\subsection{Future directions}
\label{sec:other_threats}
While all the above discussions are on practical systems that can be deployed in TEEs, most of the following paragraphs constitute exciting research possibilities rather than available solutions.

\textbf{Microservices:}
A frequent method to deploy a pipeline of operations is microservices. In such a deployment model, a result of one microservice is provided to the next, which provides its results to the next, and so forth. Microservices form a graph of operations applied to some input. For example, one microservice could be fetching some database data, the other parsing input through LLMs, and the final pushing some data to the database. Frameworks such as Marblerun\footnote{https://github.com/edgelesssys/marblerun} allow for creating proofs that all microservices are attested. However, verifying such proofs would require knowing how the microservices are structured, a competitive advantage that companies are frequently unwilling to share. 

\textbf{Wrappers on proprietary models:} Our flow focuses on private company deployments leveraging public models. However, some model providers offer their models as simple wrappers over proprietary FMs. For example, an insurance company might prepend some data to a query with additional context and submit that for evaluation to a GPT4 model. CSPs offer such models as deployments on their platforms. While there does not need to be any trust between the FM provider and the CSP like in our deployment flow, there necessarily needs to be trust between the user company and the FM provider. As we discuss in depth in the next paragraph, eliminating such trust would mean that some competitive advantage of the FM provider would need to be eliminated.

\textbf{Eliminating the need for trust in the model provider:} In our flow, users can attest that their data is processed in a TEE if they directly connect to it. This could be manifested, for example, with a special symbol in a browser like we currently do with TLS. Such a possibility is already a great improvement over the current state-of-the-art, which makes privacy-aware users notice companies that work with TEEs.

However, the user needs to trust the model provider that they are running the model they have been promised and not using their data maliciously. While in our deployment strategy, this does not matter, it would matter if the model deployed is a public service such as public LLM chats or the aforementioned wrappers over proprietary models. TEEs resolve this by attesting the running software within them. Yet, such code frequently comprises the competitive advantage of FM providers, so it is hard for them or the CSPs to release it.

We could create standard codes that could be attested and deploy certain FMs, which would be measured and hashes compared with the promised references. Such a flow would allow the users to verify the code as well as the fact that it is running in a TEE. This problem is part of a larger discussion within the community~\cite{delignat-lavaudWhyShouldTrust2023}.

\textbf{Other assurances:} The aforementioned ability to have a standardized deployment code would also enable a unique opportunity for publishing official accuracy results of models. One could imagine a marketplace in which models are sold. Leveraging TEEs allows the model provider to prove securely to any third party in such a marketplace that a given model achieves a certain accuracy on standard datasets without the third party needing access to the model. Any of these assurances would be challenging to provably obtain without using TEEs.

\section{Lifting FMs into TEEs}
Our flow and protections against a modern FM adversary generalize to all TEEs. To show a practical deployment, we select a subset of available TEEs and implement the data pipeline, show the insights we gained, and release our configuration. We limit ourselves to TEEs widely offered by major CSPs due to the practicality of such a choice. This reduces the available options to two types of TEEs from the largest CPU vendors, \intel\ and AMD. We selected \intels\ TEEs for two reasons. Firstly, they provide us with the two common ways of implementing TEEs (virtual machine and application-based) while using the same machine, allowing for an apples-to-apples comparison without any performance result scaling. Secondly, they include support of AMX, a specialized, on-chip, AI hardware accelerator introducing CPU native support for formats such as brain floating numbers and 8-bit integers, increasing overall performance. We also discuss GPU-based TEEs in Section~\ref{sec:gpu_perf}. We outline the basic properties of these TEEs and show how they can be practically leveraged. 

\label{sec:hardship}

\begin{figure}[]
  \centering
  \includegraphics[width=\linewidth]{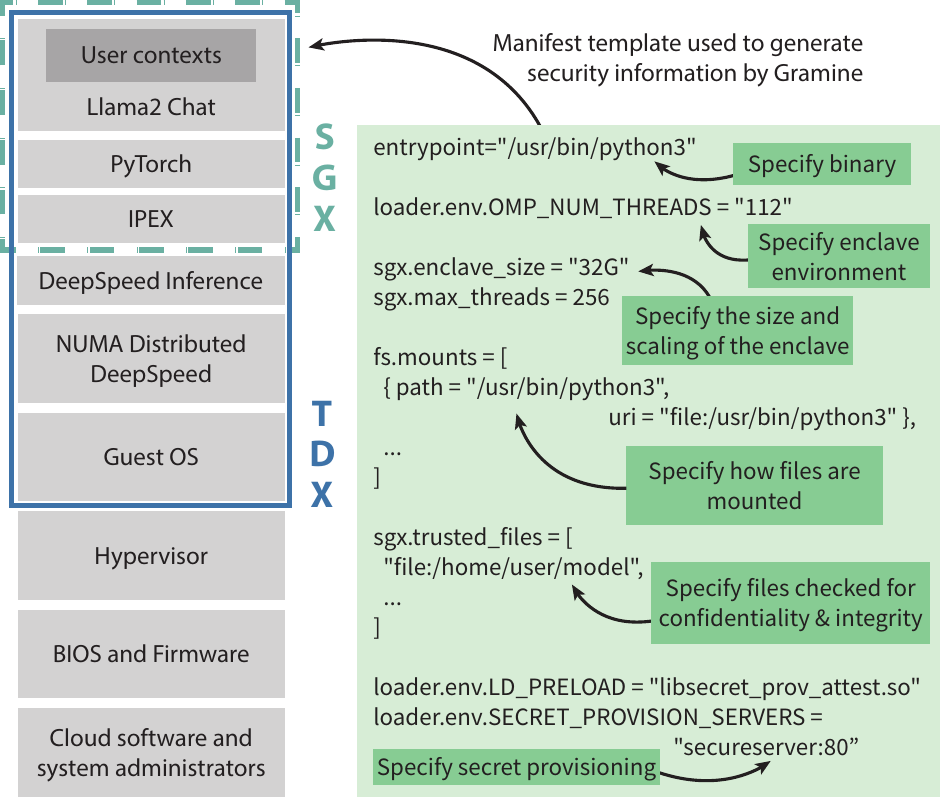}
  \caption{A differentiation between TDX and SGX with an extract from our Gramine manifest template file containing all the information needed to provide security to a Gramine-based TEE running an FM workload.}
  \label{fig:TEEs}
\end{figure}

\subsection{Software Guard Extension (SGX)}
\label{sec:SGX}

SGX is an application-based TEE. In the security model of SGX, the applications can run on bare metal without any virtualization or within a VM. The TCB involves only parts of the applications and the hardware. The SGX programming model differentiates between SGX non-protected and SGX-protected parts of the program. The protected part of the program is located within an enclave and is safeguarded by SGX capabilities, while the non-protected part is located outside of SGX. All the data in the enclave is protected by memory encryption and integrity checks. Operations requiring leaving the enclave to run the SGX non-protected part of the program (e.g., reading a file) securely save its state for later reuse and clear the caches for security reasons.

SGX enclaves are frequently deployed on top of library OSs created specifically for TEEs, such as Gramine~\cite{tsaiGrapheneSGXPracticalLibrary2017} or Occlum~\cite{shenOcclumSecureEfficient2020}. These are lightweight layers between the host system and an application intercepting any \verb|systemcalls| to ensure they are conducted securely. These addressed some inconveniences of the original SGX SDK, which required users to rewrite their applications with secure and insecure sections. This implied using the necessary manual instructions for operations such as entering and leaving sections, implementing protections for loading files, and conducting attestation. %

Gramine is a well-established, multi-company-backed, open-source project\footnote{\url{https://github.com/gramineproject/gramine}}. Figure~\ref{fig:TEEs} shows our software stack in which we use Gramine to run PyTorch and Intel Extensions for Pytorch (IPEX) within SGX without major code modifications. It simplifies using common features of TEEs by creating security in the background. For example, it would automatically implement instructions for leaving and entering the enclave. While loading files, it would also conduct integrity and encryption protections for the users and would allow for setting up attestation. While executing, Gramine intercepts and emulates application system calls. Depending on the system call, the functionality can be provided efficiently without exiting the SGX enclave. On the other hand, system calls like file accesses require an exit, which incurs additional costs. %

Gramine exposes these features via an application-dependent Manifest file. Figure~\ref{fig:TEEs} also shows parts of the Manifest we crafted for our FMs and the layers we moved into SGX. A Manifest allows the users to outline the size of the enclave, the number of threads, what should be run as an entry point binary, which files can be trusted (i.e., loaded without any integrity and confidentiality checks), which files should be allowed (i.e., which files can be accessed), where to obtain the cryptographic decryption keys, and how to conduct attestation. It then uses this information to generate the necessary cryptographic information passed to the enclave. Manifest files are created from templates with examples in the Gramine repository. We release our Manifests to ease adoption of FMs within TEEs.

\subsection{Trusted Domain Extensions (TDX)}
TDX is a virtual machine (VM) based TEE that introduces security features using a hardened hardware-enabled hypervisor. In the TDX security model, the entire VM is protected. This considerably simplifies the development as the user does not need to worry about special functions when leaving or entering the enclave. Users also do not need to find all the files their application uses and can run their programs within standard Linux OS such as Ubuntu. Furthermore, this approach fits well with the CSP virtualization trend. For our implementation, VMs enable Deepspeed within the enclave, opening venues for easier accelerator support and multi-node inference. However, the price for this comfort is considerably higher TCB as the whole VM OS must be trusted. Using VMs implies a virtualization performance tax that we will show in Section~\ref{sec:performance}, which can be similar to the overheads of SGX. 

To use TDX, one must create an appropriate VM definition file. It specifies details such as what file should be used to boot the VM, how to map the appropriate virtual cores to physical ones, and the available memory. This definition file considerably influences the VM's performance and can have a larger impact on the final performance than enabling or disabling the security features of TDX. We provide our optimized definition file. %

\subsection{Performance optimizations}
As our workload does not use many IOs on the critical path, we found that the main driver for the performance of the TEEs for FMs is twofold:
\begin{enumerate}
    \itemsep0em
    \item the ability to use special purpose accelerators inside the CPU like \intel\ Advanced Vector Extension (AVX) or Advanced Matrix Extension (AMX);
    \item efficient use of memory controllers and reaching peak memory bandwidth.
\end{enumerate}
The former can be enabled by appropriate configuration and leveraging frameworks such as IPEX. We found that once such accelerators are used, the performance overheads are dominated by the latter, which is optimized by, in general, lowering the memory pressure. Leveraging accelerators such as AMX also allowed us to achieve this goal by moving from \verb|float32| to a native hardware support of \verb|bfloat16| and \verb|int8|.

For both TDX, and SGX, we also optimized our initial performance results by using TCMalloc~\cite{durnerImpactMemoryAllocation2019}, and an Open MP~\cite{dagumOpenMPIndustryStandard1998} version suitable for \intel\ processors. The former reduces the memory pressure, while the latter makes better use of hardware. Furthermore, we found sub-NUMA clustering to have a large influence on both SGX and TDX. Sub-NUMA clustering in \intel\ CPUs splits a single CPU into multiple NUMA domains and typically improves the performance for ML workloads. Currently, the TEE drivers and the OS do not support sub-NUMA domains, resulting in inefficient memory placement. Instead, a single NUMA domain per CPU achieves higher performance. As a result, we disabled sub-NUMA clustering during our experiments.

For TDX specifically, we used huge pages~\cite{panwarMakingHugePages2018}, which reduced the number of necessary translation lookaside buffer (TLB) accesses, decreasing memory access latency. We similarly observed higher performance when not exposing a CPU core's second logical thread (hyper thread) to the VM. In its default configuration, PyTorch only executes on the first logical thread of a core, ignoring the hyperthreads, making the second thread introduce noise.

\section{Performance of LLMs in TEEs}

\begin{figure*}
  \centering
  \includegraphics[width=\linewidth]{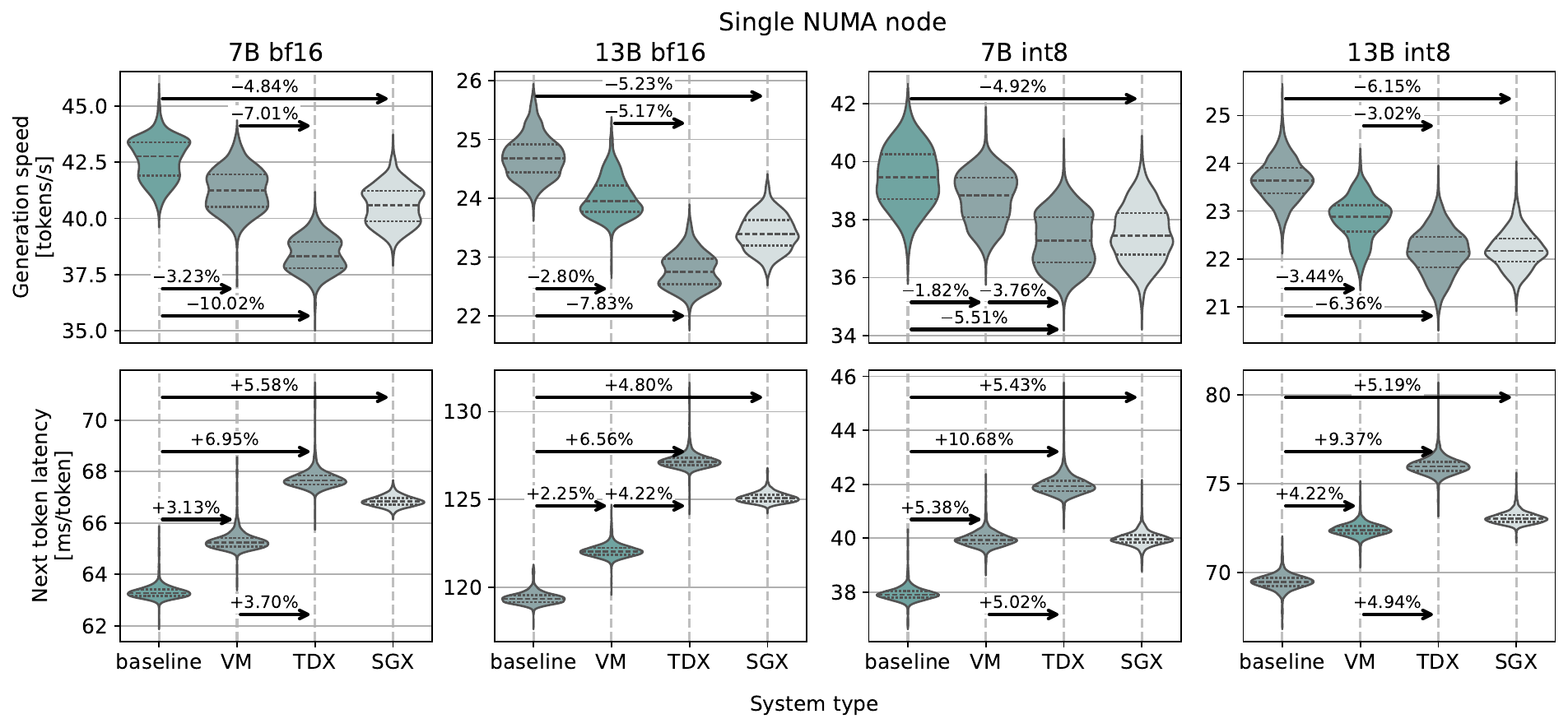}
  \caption{The TEE generation speed reductions and latency overheads are within 4-10\% for TDX and SGX.}
  \label{fig:overall}
\end{figure*}

\label{sec:performance}
We run FMs within TDX and SGX using our best performing and optimized configurations, and show the performance of the Llama2~\cite{touvronLlamaOpenFoundation2023} model family as it represents modern FMs well. We show how TEEs are a practical solution for providing security to modern models from a performance perspective with considerably lower overheads than previously reported 100s of percent~\cite{akramPerformanceAnalysisScientific2021}. This is even though LLMs are large and complex models, creating considerable memory pressure. As TEE overheads stem from either IO operations or the encryption and decryption of the accessed memory, such models maximize the cost of using TEEs. We focus on user-perceived performance: throughput and latency as measured by the number of tokens per second that our pipeline generates and the time to receive a token in the pipeline. While many metrics were created to measure the performance of ML models~\cite{dehghaniEfficiencyMisnomer2021}, our choice focuses on real-world experience.

\subsection{Experimental setup}
In our experiments, we used a dual socket \intel\ Xeon\textsuperscript{\textregistered}\ Platinum GOLD 6530 system with 32 cores and 16x32GiB 4800MHz DDR5 memory. We used Ubuntu 23.10, with Python version 3.10.14, PyTorch version 2.2.0, transformers version 4.35.2, \intel\ extension for PyTorch (IPEX) version 2.2.0, and oneCCL PyTorch bindings version 2.2.0. We measured the throughput and latency for four setups. Baseline results represent results coming from running the model on the bare metal instance, Gramine SGX represents the results of Gramine v1.7 backend running on SGX, VM the results of running a raw VM without any security features, and TDX the results of running on TDX. The machine is configured with disabled sub-NUMA clustering, and the workload is restricted to a single thread per core in the VM definition or via \verb|numactl|.

In all experiments, we used batch size six for throughput and batch size one for latency. A larger batch size means increased latency but higher throughput as less data movement is required per token for each computation within the LLM. The different inputs batched together can be computed on each layer of the LLM, and a combined result can be forwarded to the next layer. Each layer takes longer than for a single input (increased latency) but shorter than for $N$ separate inputs (increased throughput). We used 1024 input and 128 output tokens. We conducted multiple generation runs where, for each model size and system, we measured at least 1000 tokens. For latency, we measured the generation time for each token and plotted these as a distribution. For throughput, we plotted the inverse of each token generation time. We filtered out outlier tokens by excluding times with an absolute Z-score larger than 3. These constitute on average 0.64\% of the data ranging from 0.2\% to 1.2\% and do not contribute to the discussion but create considerable noise on the plots. In all experiments, we used two inference datatypes: brain floating point (\verb|bfloat16|) and the quantized integers (\verb|int8|). 

\subsection{Latency and Generation Speed}
We compare the next token latency of Llama2 inference running in different configurations. Figure~\ref{fig:overall} shows the generation speed of tokens (batch size = 6 and beam size = 4) and the next token latency (batch size = 1, beam size = 1). The overhead of Gramine-SGX is between 4.58-6.55\% while the TDX overhead is between 5.83-11.14\%. These results demonstrate that running inside a VM has a non-negligible performance overhead of 1.85-5.11\% to which TDX adds additional overhead of 3.12-7.54\%. The performance of SGX is in between a VM and TDX. The results for different data types show that \verb|int8| generally obtains similar generation speeds as \verb|bf16| but almost half the latency. However, the overheads for \verb|int8| are better in the case of generation speed and worse in the case of latency. We believe that this might be because, in the latency case, there is a constant memory access latency overhead due to memory protections that TEEs introduce, which is more pronounced in the higher-performing latency results. For the throughput, because our inference state is in \verb|int8|, there is less memory movement, which results in lower overheads due to necessary address translations from guest to host memory.

In our deployment, SGX runs on top of the host OS. The host OS has more privileges than a VM and exposes the hardware more directly to SGX. TDX, on the other hand, does not have direct access to some hardware features and needs to go through virtualization layers, such as guest address translations that are not present in SGX. As these TEEs do not access any network or drive during the computation, most of the overhead comes from the protected memory accesses. It is important to notice that most machines rented from CSPs run VMs. Thus, the visible overheads will be lower as the usual baseline is an already virtualized system and not a bare metal machine like ours. The overheads of SGX over baremetal and TDX over VM are similar. All systems have latency considerably lower than the average human reading speed of 200 ms/word (300 words/min)~\cite{raynerMuchReadLittle2016}.

\section{Discussion}
\label{sec:discussion}
\subsection{Accelerator based TEEs}
\label{sec:gpu_perf}
While an exciting future direction, accelerator TEEs are not yet widespread. At the time of writing, GPU-based TEE offered by Nvidia in the H100 series is the only natively accelerated TEE solution entering the space in scale. Furthermore, even H100s with CC enabled are available only in a single CSP (Azure)~\cite{AnnouncingAzureConfidential}. Given the relative scarcity and cost of these accelerators, CPUs with on-chip accelerators such as AMX might offer a more pragmatic alternative for users who require CC to protect their FM models, queries, and other sensitive data. Our approach applies, in principle, to TEEs with an accelerator. Our software stack, in addition with DeepSpeed, offloads to most accelerators. Accelerators such as H100 require a CPU TEE such as SEV-SNP or TDX and would be brought into the trusted computing base (TCB) leveraging attestation. 

While deploying practically these accelerators, it is important to consider their threat model, which is currently different from the CPU TEE threat model. For example, H100s do not encrypt their HBM memory~\cite{dhanuskodiCreatingFirstConfidential2023}, compared to CPUs that do. While in CPU-based systems, communication between different NUMA nodes is transparently encrypted, interconnects such as PCIe and NVLINK do not yet have such a feature~\cite{dhanuskodiCreatingFirstConfidential2023}. In existing systems, any communication with the accelerator requires using bounce buffers, where data is encrypted and copied such that the accelerator comprehends it. Such a solution introduces performance overhead for workloads with a considerable IO between the CPU and the accelerator.

To the best of our knowledge, these and other performance overheads have not yet been widely verified as of the writing of this article. Evaluations of Llama2 7B on H100s in CC mode in literature show overheads of 100-150\% of total runtime~\cite{SecuringAIInference}. Vendor-published data quantifying the communication overhead measures it at about 45\%~\cite{nertneyConfidentialComputeNVIDIA2023} for Resnet50 training on H100s with literature noting overheads of up to 90\% depending on the buffer size~\cite{SecuringAIInference}. Solutions such as TDX Connect~\cite{chengIntelTDXDemystified2024} and SEV IO~\cite{SEVTIOFirmwareInterface2023} are in progress to address communication overheads by leveraging PCIe's Integrity and Data
Encryption (IDE) and TEE Device Interface Security Protocol (TDISP)~\cite{TEEDeviceInterface}. To the best of our knowledge, there also does not exist a solution allowing direct memory access between accelerator-based TEEs without the intermediate CPU involvement, which is crucial for performantly running larger models that might not fit on a single accelerator~\cite{SecuringAIInference}.

\subsection{Process or VM TEEs}
We discuss the properties of TEEs and compare application- and VM-based TEEs. Both types provide better security than other protection mechanisms at less than 10\% performance cost. TEEs can protect the confidentiality, integrity, and privacy of the models and user data while not requiring retraining or modifying the accuracy. While both types provide these features, they differ in some aspects. 

VM-based TEEs provide a well-known and deployed abstraction. They do not require application changes. This results in a known development and deployment life cycle. However, the cost that the user pays for this is twofold: a considerably increased trusted computing base (TCB), and a performance virtualization tax. The TCB for VM-based solutions includes the whole OS, such as Ubuntu. Application-based TEEs are suitable for smaller models as programming them is more challenging and typically requires a library OS to lift and shift applications into the TEE or develop the application from scratch. However, they perform better for a single socket and provide stronger security properties. Their TCB is only a part of a library OS like Gramine, which is considerably smaller than a fully-fetched OS. While the software support for enabling the security is currently better for application-based TEEs, this is rapidly changing with projects such as Gramine adding the necessary support for VM-based TEEs, making the attestation and integrity/confidentiality verification easier.

\subsection{Training}
\label{sec:training}
While in our work, we focus on inference, introducing TEEs to the training process enables similar protections with further assurances that can be put on the training data. TEEs can verify whether a dataset conforms to some standards, such as lack of hate speech, personal information, copyright violations, or gender bias, in a securely provable way. Furthermore, this implies that the resulting model and the data used in the training process can be bound. This can be achieved using a secure hash function, combining the dataset's hash and FM hash with the certificate provided by the hardware. This method allows the training party to securely prove to any other party that a given dataset has been used during the training of a particular model.

\section{Conclusions}
We presented an emerging threat model for LLMs and discussed how modern TEEs can secure against these adversaries. We showed how TEEs can additionally provide assurances on properties such as accuracy, opening a venue for the next generation quality of service agreements. Furthermore, we have addressed the two common critiques of TEEs: ease of use and large performance overheads. We have implemented an example pipeline for LLMs within TEEs and shared our learnings and code. We have also shown in the example of a state-of-the-art Llama LLM that TEEs impose a manageable performance overhead. We see TEEs opening many exciting research venues in the future, such as enabling assurances on modern LLMs, publically trusted LLM services, and LLM marketplaces where accuracy is proven property. We believe TEEs are the practical and existing solution to the problem of private and secure LLM evaluations that provide the deployments with baseline security on which others can build.

\section*{Acknowledgements}
This research was conducted as part of the “UrbanTwin: An urban digital twin for climate
action: Assessing policies and solutions for energy, water and infrastructure” project, funded by ETH-Domain Joint Initiative program in the Strategic Area Energy, Climate and Sustainable Environment, with additional support from Intel Corporation. We thank Intel for providing hardware resources, Cory Cornelius for his valuable feedback, and Madlen Koblinger for assisting with the design of figures.

\bibliography{library}

\begin{thebibliography}{70}
\providecommand{\natexlab}[1]{#1}
\providecommand{\url}[1]{\texttt{#1}}
\expandafter\ifx\csname urlstyle\endcsname\relax
  \providecommand{\doi}[1]{doi: #1}\else
  \providecommand{\doi}{doi: \begingroup \urlstyle{rm}\Url}\fi

\bibitem[Ann()]{AnnouncingAzureConfidential}
Announcing {{Azure}} confidential {{VMs}} with {{NVIDIA H100 Tensor Core GPUs}}
  in {{Preview}}.
\newblock URL
  \url{https://techcommunity.microsoft.com/t5/azure-confidential-computing/announcing-azure-confidential-vms-with-nvidia-h100-tensor-core/ba-p/3975389}.

\bibitem[SEV()]{SEVTIOFirmwareInterface2023}
{{SEV-TIO Firmware Interface Specification}}.

\bibitem[Sec()]{SecuringAIInference}
Securing {{AI Inference}} in the {{Cloud}}: {{Is CPU-GPU Confidential Computing
  Ready}}?
\newblock URL \url{https://ieeexplore.ieee.org/document/10643934}.

\bibitem[TEE()]{TEEDeviceInterface}
{{TEE Device Interface Security Protocol}} ({{TDISP}}) | {{PCI-SIG}}.
\newblock URL
  \url{https://pcisig.com/tee-device-interface-security-protocol-tdisp}.

\bibitem[Acar et~al.()Acar, Aksu, Uluagac, and
  Conti]{acarSurveyHomomorphicEncryption2018}
Acar, A., Aksu, H., Uluagac, A.~S., and Conti, M.
\newblock A {{Survey}} on {{Homomorphic Encryption Schemes}}: {{Theory}} and
  {{Implementation}}.
\newblock 51\penalty0 (4):\penalty0 79:1--79:35.
\newblock ISSN 0360-0300.
\newblock \doi{10.1145/3214303}.
\newblock URL \url{https://dl.acm.org/doi/10.1145/3214303}.

\bibitem[Akram et~al.()Akram, Giannakou, Akella, Lowe-Power, and
  Peisert]{akramPerformanceAnalysisScientific2021}
Akram, A., Giannakou, A., Akella, V., Lowe-Power, J., and Peisert, S.
\newblock Performance {{Analysis}} of {{Scientific Computing Workloads}} on
  {{General Purpose TEEs}}.
\newblock In \emph{2021 {{IEEE International Parallel}} and {{Distributed
  Processing Symposium}} ({{IPDPS}})}, pp.\  1066--1076.
\newblock \doi{10.1109/IPDPS49936.2021.00115}.
\newblock URL \url{https://ieeexplore.ieee.org/document/9460547}.

\bibitem[Araci()]{araciFinBERTFinancialSentiment2019}
Araci, D.
\newblock {{FinBERT}}: {{Financial Sentiment Analysis}} with {{Pre-trained
  Language Models}}.
\newblock URL \url{http://arxiv.org/abs/1908.10063}.

\bibitem[Awais et~al.()Awais, Naseer, Khan, Anwer, Cholakkal, Shah, Yang, and
  Khan]{awaisFoundationalModelsDefining2023}
Awais, M., Naseer, M., Khan, S., Anwer, R.~M., Cholakkal, H., Shah, M., Yang,
  M.-H., and Khan, F.~S.
\newblock Foundational {{Models Defining}} a {{New Era}} in {{Vision}}: {{A
  Survey}} and {{Outlook}}.
\newblock URL \url{http://arxiv.org/abs/2307.13721}.

\bibitem[Ben-Nun \& Hoefler()Ben-Nun and
  Hoefler]{ben-nunDemystifyingParallelDistributed2019}
Ben-Nun, T. and Hoefler, T.
\newblock Demystifying {{Parallel}} and {{Distributed Deep Learning}}: {{An
  In-depth Concurrency Analysis}}.
\newblock 52\penalty0 (4):\penalty0 65:1--65:43.
\newblock ISSN 0360-0300.
\newblock \doi{10.1145/3320060}.
\newblock URL \url{https://doi.org/10.1145/3320060}.

\bibitem[Birkholz et~al.()Birkholz, Thaler, Richardson, Smith, and
  Pan]{birkholzRemoteATtestationProcedureS2023}
Birkholz, H., Thaler, D., Richardson, M., Smith, N., and Pan, W.
\newblock Remote {{ATtestation procedureS}} ({{RATS}}) {{Architecture}}.
\newblock URL \url{https://datatracker.ietf.org/doc/rfc9334}.

\bibitem[Boenisch()]{boenischSystematicReviewModel2021}
Boenisch, F.
\newblock A {{Systematic Review}} on {{Model Watermarking}} for {{Neural
  Networks}}.
\newblock 4:\penalty0 729663.
\newblock ISSN 2624-909X.
\newblock \doi{10.3389/fdata.2021.729663}.
\newblock URL \url{http://arxiv.org/abs/2009.12153}.

\bibitem[Brown et~al.()Brown, Mann, Ryder, Subbiah, Kaplan, Dhariwal,
  Neelakantan, Shyam, Sastry, Askell, Agarwal, Herbert-Voss, Krueger, Henighan,
  Child, Ramesh, Ziegler, Wu, Winter, Hesse, Chen, Sigler, Litwin, Gray, Chess,
  Clark, Berner, McCandlish, Radford, Sutskever, and
  Amodei]{brownLanguageModelsAre2020}
Brown, T.~B., Mann, B., Ryder, N., Subbiah, M., Kaplan, J., Dhariwal, P.,
  Neelakantan, A., Shyam, P., Sastry, G., Askell, A., Agarwal, S.,
  Herbert-Voss, A., Krueger, G., Henighan, T., Child, R., Ramesh, A., Ziegler,
  D.~M., Wu, J., Winter, C., Hesse, C., Chen, M., Sigler, E., Litwin, M., Gray,
  S., Chess, B., Clark, J., Berner, C., McCandlish, S., Radford, A., Sutskever,
  I., and Amodei, D.
\newblock Language {{Models}} are {{Few-Shot Learners}}.
\newblock URL \url{http://arxiv.org/abs/2005.14165}.

\bibitem[Burkhalter et~al.()Burkhalter, Hithnawi, Viand, Shafagh, and
  Ratnasamy]{burkhalterTimeCryptEncryptedData2020}
Burkhalter, L., Hithnawi, A., Viand, A., Shafagh, H., and Ratnasamy, S.
\newblock \{\vphantom\}{{TimeCrypt}}\vphantom\{\}: {{Encrypted Data Stream
  Processing}} at {{Scale}} with {{Cryptographic Access Control}}.
\newblock pp.\  835--850.
\newblock ISBN 978-1-939133-13-7.
\newblock URL
  \url{https://www.usenix.org/conference/nsdi20/presentation/burkhalter}.

\bibitem[Carlini et~al.()Carlini, Tramèr, Wallace, Jagielski, Herbert-Voss,
  Lee, Roberts, Brown, Song, Erlingsson, Oprea, and
  Raffel]{carliniExtractingTrainingData2021}
Carlini, N., Tramèr, F., Wallace, E., Jagielski, M., Herbert-Voss, A., Lee,
  K., Roberts, A., Brown, T., Song, D., Erlingsson, U., Oprea, A., and Raffel,
  C.
\newblock Extracting {{Training Data}} from {{Large Language Models}}.
\newblock pp.\  2633--2650.
\newblock ISBN 978-1-939133-24-3.
\newblock URL
  \url{https://www.usenix.org/conference/usenixsecurity21/presentation/carlini-extracting}.

\bibitem[Chen et~al.()Chen, Zaharia, and Zou]{chenHowChatGPTBehavior2023}
Chen, L., Zaharia, M., and Zou, J.
\newblock How is {{ChatGPT}}'s behavior changing over time?
\newblock URL \url{http://arxiv.org/abs/2307.09009}.

\bibitem[Cheng et~al.({\natexlab{a}})Cheng, Ozga, Valdez, Ahmed, Gu, Jamjoom,
  Franke, and Bottomley]{chengIntelTDXDemystified2023}
Cheng, P.-C., Ozga, W., Valdez, E., Ahmed, S., Gu, Z., Jamjoom, H., Franke, H.,
  and Bottomley, J.
\newblock Intel {{TDX Demystified}}: {{A Top-Down Approach}}, {\natexlab{a}}.
\newblock URL \url{http://arxiv.org/abs/2303.15540}.

\bibitem[Cheng et~al.({\natexlab{b}})Cheng, Ozga, Valdez, Ahmed, Gu, Jamjoom,
  Franke, and Bottomley]{chengIntelTDXDemystified2024}
Cheng, P.-C., Ozga, W., Valdez, E., Ahmed, S., Gu, Z., Jamjoom, H., Franke, H.,
  and Bottomley, J.
\newblock Intel {{TDX Demystified}}: {{A Top-Down Approach}}.
\newblock {\natexlab{b}}.
\newblock ISSN 0360-0300.
\newblock \doi{10.1145/3652597}.
\newblock URL \url{https://dl.acm.org/doi/10.1145/3652597}.

\bibitem[Chrapek et~al.()Chrapek, Khalilov, and
  Hoefler]{chrapekHEARHomomorphicallyEncrypted2023}
Chrapek, M., Khalilov, M., and Hoefler, T.
\newblock {{HEAR}}: {{Homomorphically Encrypted Allreduce}}.
\newblock In \emph{Proceedings of the {{International Conference}} for {{High
  Performance Computing}}, {{Networking}}, {{Storage}} and {{Analysis}}},
  {{SC}} '23, pp.\  1--17. Association for Computing Machinery.
\newblock ISBN 9798400701092.
\newblock \doi{10.1145/3581784.3607099}.
\newblock URL \url{https://dl.acm.org/doi/10.1145/3581784.3607099}.

\bibitem[Costan \& Devadas()Costan and Devadas]{costanIntelSGXExplained2016}
Costan, V. and Devadas, S.
\newblock Intel {{SGX Explained}}.
\newblock URL \url{https://eprint.iacr.org/2016/086}.

\bibitem[Cui et~al.()Cui, Li, Yan, Chen, and
  Yuan]{cuiChatLawOpenSourceLegal2023}
Cui, J., Li, Z., Yan, Y., Chen, B., and Yuan, L.
\newblock {{ChatLaw}}: {{Open-Source Legal Large Language Model}} with
  {{Integrated External Knowledge Bases}}.
\newblock URL \url{http://arxiv.org/abs/2306.16092}.

\bibitem[Dagum \& Menon()Dagum and Menon]{dagumOpenMPIndustryStandard1998}
Dagum, L. and Menon, R.
\newblock {{OpenMP}}: An industry standard {{API}} for shared-memory
  programming.
\newblock 5\penalty0 (1):\penalty0 46--55.
\newblock ISSN 1558-190X.
\newblock \doi{10.1109/99.660313}.
\newblock URL \url{https://ieeexplore.ieee.org/abstract/document/660313}.

\bibitem[Dehghani et~al.()Dehghani, Tay, Arnab, Beyer, and
  Vaswani]{dehghaniEfficiencyMisnomer2021}
Dehghani, M., Tay, Y., Arnab, A., Beyer, L., and Vaswani, A.
\newblock The {{Efficiency Misnomer}}.
\newblock URL \url{https://openreview.net/forum?id=iulEMLYh1uR}.

\bibitem[Delignat-Lavaud et~al.()Delignat-Lavaud, Fournet, Vaswani, Clebsch,
  Riechert, Costa, and Russinovich]{delignat-lavaudWhyShouldTrust2023}
Delignat-Lavaud, A., Fournet, C., Vaswani, K., Clebsch, S., Riechert, M.,
  Costa, M., and Russinovich, M.
\newblock Why {{Should I Trust Your Code}}? {{Confidential}} computing enables
  users to authenticate code running in {{TEEs}}, but users also need evidence
  this code is trustworthy.
\newblock 21\penalty0 (4):\penalty0 Pages 30:94--Pages 30:122.
\newblock ISSN 1542-7730.
\newblock \doi{10.1145/3623460}.
\newblock URL \url{https://dl.acm.org/doi/10.1145/3623460}.

\bibitem[Dhanuskodi et~al.()Dhanuskodi, Guha, Krishnan, Manjunatha, O'Connor,
  Nertney, and Rogers]{dhanuskodiCreatingFirstConfidential2023}
Dhanuskodi, G., Guha, S., Krishnan, V., Manjunatha, A., O'Connor, M., Nertney,
  R., and Rogers, P.
\newblock Creating the {{First Confidential GPUs}}: {{The}} team at {{NVIDIA}}
  brings confidentiality and integrity to user code and data for accelerated
  computing.
\newblock 21\penalty0 (4):\penalty0 Pages 40:68--Pages 40:93.
\newblock ISSN 1542-7730.
\newblock \doi{10.1145/3623393.3623391}.
\newblock URL \url{https://dl.acm.org/doi/10.1145/3623393.3623391}.

\bibitem[Dowlin et~al.()Dowlin, Gilad-Bachrach, Laine, Lauter, Naehrig, and
  Wernsing]{dowlinCryptoNetsApplyingNeural}
Dowlin, N., Gilad-Bachrach, R., Laine, K., Lauter, K., Naehrig, M., and
  Wernsing, J.
\newblock {{CryptoNets}}: {{Applying Neural Networks}} to {{Encrypted Data}}
  with {{High Throughput}} and {{Accuracy}}.

\bibitem[Durner et~al.()Durner, Leis, and
  Neumann]{durnerImpactMemoryAllocation2019}
Durner, D., Leis, V., and Neumann, T.
\newblock On the {{Impact}} of {{Memory Allocation}} on {{High-Performance
  Query Processing}}.
\newblock In \emph{Proceedings of the 15th {{International Workshop}} on {{Data
  Management}} on {{New Hardware}}}, {{DaMoN}}'19, pp.\  1--3. Association for
  Computing Machinery.
\newblock ISBN 978-1-4503-6801-8.
\newblock \doi{10.1145/3329785.3329918}.
\newblock URL \url{https://dl.acm.org/doi/10.1145/3329785.3329918}.

\bibitem[Fan et~al.()Fan, Ng, and Chan]{fanRethinkingDeepNeural2019}
Fan, L., Ng, K.~W., and Chan, C.~S.
\newblock Rethinking {{Deep Neural Network Ownership Verification}}:
  {{Embedding Passports}} to {{Defeat Ambiguity Attacks}}.
\newblock In \emph{Advances in {{Neural Information Processing Systems}}},
  volume~32. Curran Associates, Inc.
\newblock URL
  \url{https://proceedings.neurips.cc/paper_files/paper/2019/hash/75455e062929d32a333868084286bb68-Abstract.html}.

\bibitem[Goldfarb \& Tucker()Goldfarb and
  Tucker]{goldfarbShiftsPrivacyConcerns2012}
Goldfarb, A. and Tucker, C.
\newblock Shifts in {{Privacy Concerns}}.
\newblock 102\penalty0 (3):\penalty0 349--353.
\newblock ISSN 0002-8282.
\newblock \doi{10.1257/aer.102.3.349}.
\newblock URL \url{https://www.aeaweb.org/articles?id=10.1257/aer.102.3.349}.

\bibitem[Hoekstra et~al.()Hoekstra, Lal, Pappachan, Phegade, and
  Del~Cuvillo]{hoekstraUsingInnovativeInstructions2013}
Hoekstra, M., Lal, R., Pappachan, P., Phegade, V., and Del~Cuvillo, J.
\newblock Using innovative instructions to create trustworthy software
  solutions.
\newblock In \emph{Proceedings of the 2nd {{International Workshop}} on
  {{Hardware}} and {{Architectural Support}} for {{Security}} and {{Privacy}}},
  {{HASP}} '13, pp.\ ~1. Association for Computing Machinery.
\newblock ISBN 978-1-4503-2118-1.
\newblock \doi{10.1145/2487726.2488370}.
\newblock URL \url{https://doi.org/10.1145/2487726.2488370}.

\bibitem[Kaplan()]{kaplanAMDSEVSNPStrengthening}
Kaplan, D.
\newblock {{AMD SEV-SNP}}: {{Strengthening VM Isolation}} with {{Integrity
  Protection}} and {{More}}.

\bibitem[Knauth et~al.()Knauth, Steiner, Chakrabarti, Lei, Xing, and
  Vij]{knauthIntegratingRemoteAttestation2019}
Knauth, T., Steiner, M., Chakrabarti, S., Lei, L., Xing, C., and Vij, M.
\newblock Integrating {{Remote Attestation}} with {{Transport Layer Security}}.
\newblock URL \url{http://arxiv.org/abs/1801.05863}.

\bibitem[Knott et~al.()Knott, Venkataraman, Hannun, Sengupta, Ibrahim, and
  family=Maaten]{knottCrypTenSecureMultiParty2022}
Knott, B., Venkataraman, S., Hannun, A., Sengupta, S., Ibrahim, M., and
  family=Maaten, given=Laurens, p. d.~u.
\newblock {{CrypTen}}: {{Secure Multi-Party Computation Meets Machine
  Learning}}.
\newblock URL \url{http://arxiv.org/abs/2109.00984}.

\bibitem[Kocmi \& Federmann()Kocmi and Federmann]{kocmiLargeLanguageModels2023}
Kocmi, T. and Federmann, C.
\newblock Large {{Language Models Are State-of-the-Art Evaluators}} of
  {{Translation Quality}}.
\newblock URL \url{http://arxiv.org/abs/2302.14520}.

\bibitem[Lao et~al.()Lao, Zhao, Yang, and Li]{laoDeepAuthDNNAuthentication2022}
Lao, Y., Zhao, W., Yang, P., and Li, P.
\newblock {{DeepAuth}}: {{A DNN Authentication Framework}} by {{Model-Unique}}
  and {{Fragile Signature Embedding}}.
\newblock 36\penalty0 (9):\penalty0 9595--9603.
\newblock ISSN 2374-3468.
\newblock \doi{10.1609/aaai.v36i9.21193}.
\newblock URL \url{https://ojs.aaai.org/index.php/AAAI/article/view/21193}.

\bibitem[Lee et~al.({\natexlab{a}})Lee, Kuvaiskii, Vahldiek-Oberwagner, and
  Vij]{leePrivacyPreservingMachineLearning2020}
Lee, D., Kuvaiskii, D., Vahldiek-Oberwagner, A., and Vij, M.
\newblock Privacy-{{Preserving Machine Learning}} in {{Untrusted Clouds Made
  Simple}}, {\natexlab{a}}.
\newblock URL \url{http://arxiv.org/abs/2009.04390}.

\bibitem[Lee et~al.({\natexlab{b}})Lee, Kang, Lee, Choi, Eom, Deryabin, Lee,
  Lee, Yoo, Kim, and No]{leePrivacyPreservingMachineLearning2022}
Lee, J.-W., Kang, H., Lee, Y., Choi, W., Eom, J., Deryabin, M., Lee, E., Lee,
  J., Yoo, D., Kim, Y.-S., and No, J.-S.
\newblock Privacy-{{Preserving Machine Learning With Fully Homomorphic
  Encryption}} for {{Deep Neural Network}}.
\newblock 10:\penalty0 30039--30054, {\natexlab{b}}.
\newblock ISSN 2169-3536.
\newblock \doi{10.1109/ACCESS.2022.3159694}.
\newblock URL \url{https://ieeexplore.ieee.org/abstract/document/9734024}.

\bibitem[Lewis et~al.()Lewis, Perez, Piktus, Petroni, Karpukhin, Goyal,
  Küttler, Lewis, Yih, Rocktäschel, Riedel, and
  Kiela]{lewisRetrievalAugmentedGenerationKnowledgeIntensive2020}
Lewis, P., Perez, E., Piktus, A., Petroni, F., Karpukhin, V., Goyal, N.,
  Küttler, H., Lewis, M., Yih, W.-t., Rocktäschel, T., Riedel, S., and Kiela,
  D.
\newblock Retrieval-{{Augmented Generation}} for {{Knowledge-Intensive NLP
  Tasks}}.
\newblock In \emph{Advances in {{Neural Information Processing Systems}}},
  volume~33, pp.\  9459--9474. Curran Associates, Inc.
\newblock URL
  \url{https://proceedings.neurips.cc/paper/2020/hash/6b493230205f780e1bc26945df7481e5-Abstract.html}.

\bibitem[Li et~al.()Li, Li, Dall, Gu, Nieh, Sait, and
  Stockwell]{liDesignVerificationArm2022}
Li, X., Li, X., Dall, C., Gu, R., Nieh, J., Sait, Y., and Stockwell, G.
\newblock Design and {{Verification}} of the {{Arm Confidential Compute
  Architecture}}.
\newblock pp.\  465--484.
\newblock ISBN 978-1-939133-28-1.
\newblock URL \url{https://www.usenix.org/conference/osdi22/presentation/li}.

\bibitem[McKeen et~al.()McKeen, Alexandrovich, Berenzon, Rozas, Shafi,
  Shanbhogue, and Savagaonkar]{mckeenInnovativeInstructionsSoftware2013}
McKeen, F., Alexandrovich, I., Berenzon, A., Rozas, C.~V., Shafi, H.,
  Shanbhogue, V., and Savagaonkar, U.~R.
\newblock Innovative instructions and software model for isolated execution.
\newblock In \emph{Proceedings of the 2nd {{International Workshop}} on
  {{Hardware}} and {{Architectural Support}} for {{Security}} and {{Privacy}}},
  {{HASP}} '13, pp.\ ~1. Association for Computing Machinery.
\newblock ISBN 978-1-4503-2118-1.
\newblock \doi{10.1145/2487726.2488368}.
\newblock URL \url{https://doi.org/10.1145/2487726.2488368}.

\bibitem[Mo et~al.()Mo, Tarkhani, and
  Haddadi]{moMachineLearningConfidential2023}
Mo, F., Tarkhani, Z., and Haddadi, H.
\newblock Machine {{Learning}} with {{Confidential Computing}}: {{A
  Systematization}} of {{Knowledge}}.
\newblock URL \url{http://arxiv.org/abs/2208.10134}.

\bibitem[Mulligan et~al.()Mulligan, Petri, Spinale, Stockwell, and
  Vincent]{mulliganConfidentialComputingBrave2021}
Mulligan, D.~P., Petri, G., Spinale, N., Stockwell, G., and Vincent, H. J.~M.
\newblock Confidential {{Computing}}—a brave new world.
\newblock In \emph{2021 {{International Symposium}} on {{Secure}} and {{Private
  Execution Environment Design}} ({{SEED}})}, pp.\  132--138.
\newblock \doi{10.1109/SEED51797.2021.00025}.
\newblock URL \url{https://ieeexplore.ieee.org/document/9604800}.

\bibitem[Nertney()]{nertneyConfidentialComputeNVIDIA2023}
Nertney, R.
\newblock Confidential {{Compute}} on {{NVIDIA Hopper H100}} - {{Whitepaper}}.

\bibitem[Ng()]{ngAdobeSaysIt}
Ng, T.
\newblock Adobe {{Says It Won}}’t {{Train AI Using Artists}}’ {{Work}}.
  {{Creatives Aren}}’t {{Convinced}}.
\newblock ISSN 1059-1028.
\newblock URL
  \url{https://www.wired.com/story/adobe-says-it-wont-train-ai-using-artists-work-creatives-arent-convinced/}.

\bibitem[OpenAI et~al.()OpenAI, Achiam, Adler, Agarwal, Ahmad, Akkaya, Aleman,
  Almeida, Altenschmidt, Altman, Anadkat, Avila, Babuschkin, Balaji, Balcom,
  Baltescu, Bao, Bavarian, Belgum, Bello, Berdine, Bernadett-Shapiro, Berner,
  Bogdonoff, Boiko, Boyd, Brakman, Brockman, Brooks, Brundage, Button, Cai,
  Campbell, Cann, Carey, Carlson, Carmichael, Chan, Chang, Chantzis, Chen,
  Chen, Chen, Chen, Chen, Chess, Cho, Chu, Chung, Cummings, Currier, Dai,
  Decareaux, Degry, Deutsch, Deville, Dhar, Dohan, Dowling, Dunning, Ecoffet,
  Eleti, Eloundou, Farhi, Fedus, Felix, Fishman, Forte, Fulford, Gao, Georges,
  Gibson, Goel, Gogineni, Goh, Gontijo-Lopes, Gordon, Grafstein, Gray, Greene,
  Gross, Gu, Guo, Hallacy, Han, Harris, He, Heaton, Heidecke, Hesse, Hickey,
  Hickey, Hoeschele, Houghton, Hsu, Hu, Hu, Huizinga, Jain, Jain, Jang, Jiang,
  Jiang, Jin, Jin, Jomoto, Jonn, Jun, Kaftan, Kaiser, Kamali, Kanitscheider,
  Keskar, Khan, Kilpatrick, Kim, Kim, Kim, Kirchner, Kiros, Knight, Kokotajlo,
  Kondraciuk, Kondrich, Konstantinidis, Kosic, Krueger, Kuo, Lampe, Lan, Lee,
  Leike, Leung, Levy, Li, Lim, Lin, Lin, Litwin, Lopez, Lowe, Lue, Makanju,
  Malfacini, Manning, Markov, Markovski, Martin, Mayer, Mayne, McGrew,
  McKinney, McLeavey, McMillan, McNeil, Medina, Mehta, Menick, Metz,
  Mishchenko, Mishkin, Monaco, Morikawa, Mossing, Mu, Murati, Murk, Mély,
  Nair, Nakano, Nayak, Neelakantan, Ngo, Noh, Ouyang, O'Keefe, Pachocki, Paino,
  Palermo, Pantuliano, Parascandolo, Parish, Parparita, Passos, Pavlov, Peng,
  Perelman, Peres, Petrov, Pinto, Michael, Pokorny, Pokrass, Pong, Powell,
  Power, Power, Proehl, Puri, Radford, Rae, Ramesh, Raymond, Real, Rimbach,
  Ross, Rotsted, Roussez, Ryder, Saltarelli, Sanders, Santurkar, Sastry,
  Schmidt, Schnurr, Schulman, Selsam, Sheppard, Sherbakov, Shieh, Shoker,
  Shyam, Sidor, Sigler, Simens, Sitkin, Slama, Sohl, Sokolowsky, Song,
  Staudacher, Such, Summers, Sutskever, Tang, Tezak, Thompson, Tillet,
  Tootoonchian, Tseng, Tuggle, Turley, Tworek, Uribe, Vallone, Vijayvergiya,
  Voss, Wainwright, Wang, Wang, Wang, Ward, Wei, Weinmann, Welihinda, Welinder,
  Weng, Weng, Wiethoff, Willner, Winter, Wolrich, Wong, Workman, Wu, Wu, Wu,
  Xiao, Xu, Yoo, Yu, Yuan, Zaremba, Zellers, Zhang, Zhang, Zhao, Zheng, Zhuang,
  Zhuk, and Zoph]{openaiGPT4TechnicalReport2023a}
OpenAI, Achiam, J., Adler, S., Agarwal, S., Ahmad, L., Akkaya, I., Aleman,
  F.~L., Almeida, D., Altenschmidt, J., Altman, S., Anadkat, S., Avila, R.,
  Babuschkin, I., Balaji, S., Balcom, V., Baltescu, P., Bao, H., Bavarian, M.,
  Belgum, J., Bello, I., Berdine, J., Bernadett-Shapiro, G., Berner, C.,
  Bogdonoff, L., Boiko, O., Boyd, M., Brakman, A.-L., Brockman, G., Brooks, T.,
  Brundage, M., Button, K., Cai, T., Campbell, R., Cann, A., Carey, B.,
  Carlson, C., Carmichael, R., Chan, B., Chang, C., Chantzis, F., Chen, D.,
  Chen, S., Chen, R., Chen, J., Chen, M., Chess, B., Cho, C., Chu, C., Chung,
  H.~W., Cummings, D., Currier, J., Dai, Y., Decareaux, C., Degry, T., Deutsch,
  N., Deville, D., Dhar, A., Dohan, D., Dowling, S., Dunning, S., Ecoffet, A.,
  Eleti, A., Eloundou, T., Farhi, D., Fedus, L., Felix, N., Fishman, S.~P.,
  Forte, J., Fulford, I., Gao, L., Georges, E., Gibson, C., Goel, V., Gogineni,
  T., Goh, G., Gontijo-Lopes, R., Gordon, J., Grafstein, M., Gray, S., Greene,
  R., Gross, J., Gu, S.~S., Guo, Y., Hallacy, C., Han, J., Harris, J., He, Y.,
  Heaton, M., Heidecke, J., Hesse, C., Hickey, A., Hickey, W., Hoeschele, P.,
  Houghton, B., Hsu, K., Hu, S., Hu, X., Huizinga, J., Jain, S., Jain, S.,
  Jang, J., Jiang, A., Jiang, R., Jin, H., Jin, D., Jomoto, S., Jonn, B., Jun,
  H., Kaftan, T., Kaiser, L., Kamali, A., Kanitscheider, I., Keskar, N.~S.,
  Khan, T., Kilpatrick, L., Kim, J.~W., Kim, C., Kim, Y., Kirchner, H., Kiros,
  J., Knight, M., Kokotajlo, D., Kondraciuk, L., Kondrich, A., Konstantinidis,
  A., Kosic, K., Krueger, G., Kuo, V., Lampe, M., Lan, I., Lee, T., Leike, J.,
  Leung, J., Levy, D., Li, C.~M., Lim, R., Lin, M., Lin, S., Litwin, M., Lopez,
  T., Lowe, R., Lue, P., Makanju, A., Malfacini, K., Manning, S., Markov, T.,
  Markovski, Y., Martin, B., Mayer, K., Mayne, A., McGrew, B., McKinney, S.~M.,
  McLeavey, C., McMillan, P., McNeil, J., Medina, D., Mehta, A., Menick, J.,
  Metz, L., Mishchenko, A., Mishkin, P., Monaco, V., Morikawa, E., Mossing, D.,
  Mu, T., Murati, M., Murk, O., Mély, D., Nair, A., Nakano, R., Nayak, R.,
  Neelakantan, A., Ngo, R., Noh, H., Ouyang, L., O'Keefe, C., Pachocki, J.,
  Paino, A., Palermo, J., Pantuliano, A., Parascandolo, G., Parish, J.,
  Parparita, E., Passos, A., Pavlov, M., Peng, A., Perelman, A., Peres, F. d.
  A.~B., Petrov, M., Pinto, H. P. d.~O., Michael, Pokorny, Pokrass, M., Pong,
  V., Powell, T., Power, A., Power, B., Proehl, E., Puri, R., Radford, A., Rae,
  J., Ramesh, A., Raymond, C., Real, F., Rimbach, K., Ross, C., Rotsted, B.,
  Roussez, H., Ryder, N., Saltarelli, M., Sanders, T., Santurkar, S., Sastry,
  G., Schmidt, H., Schnurr, D., Schulman, J., Selsam, D., Sheppard, K.,
  Sherbakov, T., Shieh, J., Shoker, S., Shyam, P., Sidor, S., Sigler, E.,
  Simens, M., Sitkin, J., Slama, K., Sohl, I., Sokolowsky, B., Song, Y.,
  Staudacher, N., Such, F.~P., Summers, N., Sutskever, I., Tang, J., Tezak, N.,
  Thompson, M., Tillet, P., Tootoonchian, A., Tseng, E., Tuggle, P., Turley,
  N., Tworek, J., Uribe, J. F.~C., Vallone, A., Vijayvergiya, A., Voss, C.,
  Wainwright, C., Wang, J.~J., Wang, A., Wang, B., Ward, J., Wei, J., Weinmann,
  C.~J., Welihinda, A., Welinder, P., Weng, J., Weng, L., Wiethoff, M.,
  Willner, D., Winter, C., Wolrich, S., Wong, H., Workman, L., Wu, S., Wu, J.,
  Wu, M., Xiao, K., Xu, T., Yoo, S., Yu, K., Yuan, Q., Zaremba, W., Zellers,
  R., Zhang, C., Zhang, M., Zhao, S., Zheng, T., Zhuang, J., Zhuk, W., and
  Zoph, B.
\newblock {{GPT-4 Technical Report}}.
\newblock URL \url{http://arxiv.org/abs/2303.08774}.

\bibitem[Panwar et~al.()Panwar, Prasad, and
  Gopinath]{panwarMakingHugePages2018}
Panwar, A., Prasad, A., and Gopinath, K.
\newblock Making {{Huge Pages Actually Useful}}.
\newblock In \emph{Proceedings of the {{Twenty-Third International Conference}}
  on {{Architectural Support}} for {{Programming Languages}} and {{Operating
  Systems}}}, {{ASPLOS}} '18, pp.\  679--692. Association for Computing
  Machinery.
\newblock ISBN 978-1-4503-4911-6.
\newblock \doi{10.1145/3173162.3173203}.
\newblock URL \url{https://dl.acm.org/doi/10.1145/3173162.3173203}.

\bibitem[Patil et~al.()Patil, Hase, and
  Bansal]{patilCanSensitiveInformation2023}
Patil, V., Hase, P., and Bansal, M.
\newblock Can {{Sensitive Information Be Deleted From LLMs}}? {{Objectives}}
  for {{Defending Against Extraction Attacks}}.
\newblock URL \url{https://openreview.net/forum?id=7erlRDoaV8}.

\bibitem[Petrescu \& Krishen()Petrescu and
  Krishen]{petrescuAnalyzingAnalyticsData2018}
Petrescu, M. and Krishen, A.~S.
\newblock Analyzing the analytics: Data privacy concerns.
\newblock 6\penalty0 (2):\penalty0 41--43.
\newblock ISSN 2050-3326.
\newblock \doi{10.1057/s41270-018-0034-x}.
\newblock URL \url{https://doi.org/10.1057/s41270-018-0034-x}.

\bibitem[Pinto \& Santos()Pinto and Santos]{pintoDemystifyingArmTrustZone2019}
Pinto, S. and Santos, N.
\newblock Demystifying {{Arm TrustZone}}: {{A Comprehensive Survey}}.
\newblock 51\penalty0 (6):\penalty0 130:1--130:36.
\newblock ISSN 0360-0300.
\newblock \doi{10.1145/3291047}.
\newblock URL \url{https://dl.acm.org/doi/10.1145/3291047}.

\bibitem[Rayner et~al.()Rayner, Schotter, Masson, Potter, and
  Treiman]{raynerMuchReadLittle2016}
Rayner, K., Schotter, E.~R., Masson, M. E.~J., Potter, M.~C., and Treiman, R.
\newblock So {{Much}} to {{Read}}, {{So Little Time}}: {{How Do We Read}}, and
  {{Can Speed Reading Help}}?
\newblock 17\penalty0 (1):\penalty0 4--34.
\newblock ISSN 1529-1006.
\newblock \doi{10.1177/1529100615623267}.
\newblock URL \url{https://doi.org/10.1177/1529100615623267}.

\bibitem[Ribeiro et~al.()Ribeiro, Grolinger, and
  Capretz]{ribeiroMLaaSMachineLearning2015}
Ribeiro, M., Grolinger, K., and Capretz, M.~A.
\newblock {{MLaaS}}: {{Machine Learning}} as a {{Service}}.
\newblock In \emph{2015 {{IEEE}} 14th {{International Conference}} on {{Machine
  Learning}} and {{Applications}} ({{ICMLA}})}, pp.\  896--902.
\newblock \doi{10.1109/ICMLA.2015.152}.
\newblock URL \url{https://ieeexplore.ieee.org/abstract/document/7424435}.

\bibitem[Sabt et~al.()Sabt, Achemlal, and
  Bouabdallah]{sabtTrustedExecutionEnvironment2015}
Sabt, M., Achemlal, M., and Bouabdallah, A.
\newblock Trusted {{Execution Environment}}: {{What It}} is, and {{What It}} is
  {{Not}}.
\newblock In \emph{2015 {{IEEE Trustcom}}/{{BigDataSE}}/{{ISPA}}}, volume~1,
  pp.\  57--64.
\newblock \doi{10.1109/Trustcom.2015.357}.
\newblock URL \url{https://ieeexplore.ieee.org/abstract/document/7345265}.

\bibitem[Sallam()]{sallamChatGPTUtilityHealthcare2023}
Sallam, M.
\newblock {{ChatGPT Utility}} in {{Healthcare Education}}, {{Research}}, and
  {{Practice}}: {{Systematic Review}} on the {{Promising Perspectives}} and
  {{Valid Concerns}}.
\newblock 11\penalty0 (6):\penalty0 887.
\newblock ISSN 2227-9032.
\newblock \doi{10.3390/healthcare11060887}.
\newblock URL \url{https://www.mdpi.com/2227-9032/11/6/887}.

\bibitem[Schneider et~al.()Schneider, Masti, Shinde, Capkun, and
  Perez]{schneiderSoKHardwaresupportedTrusted2022}
Schneider, M., Masti, R.~J., Shinde, S., Capkun, S., and Perez, R.
\newblock {{SoK}}: {{Hardware-supported Trusted Execution Environments}}.
\newblock URL \url{http://arxiv.org/abs/2205.12742}.

\bibitem[Sharir et~al.()Sharir, Peleg, and Shoham]{sharirCostTrainingNLP2020}
Sharir, O., Peleg, B., and Shoham, Y.
\newblock The {{Cost}} of {{Training NLP Models}}: {{A Concise Overview}}.
\newblock URL \url{http://arxiv.org/abs/2004.08900}.

\bibitem[Shen et~al.()Shen, Tian, Chen, Chen, Wang, Xu, Xia, and
  Yan]{shenOcclumSecureEfficient2020}
Shen, Y., Tian, H., Chen, Y., Chen, K., Wang, R., Xu, Y., Xia, Y., and Yan, S.
\newblock Occlum: {{Secure}} and {{Efficient Multitasking Inside}} a {{Single
  Enclave}} of {{Intel SGX}}.
\newblock In \emph{Proceedings of the {{Twenty-Fifth International Conference}}
  on {{Architectural Support}} for {{Programming Languages}} and {{Operating
  Systems}}}, {{ASPLOS}} '20, pp.\  955--970. Association for Computing
  Machinery.
\newblock ISBN 978-1-4503-7102-5.
\newblock \doi{10.1145/3373376.3378469}.
\newblock URL \url{https://dl.acm.org/doi/10.1145/3373376.3378469}.

\bibitem[Szyller \& Asokan()Szyller and
  Asokan]{szyllerConflictingInteractionsProtection2023}
Szyller, S. and Asokan, N.
\newblock Conflicting interactions among protection mechanisms for machine
  learning models.
\newblock In \emph{Proceedings of the {{Thirty-Seventh AAAI Conference}} on
  {{Artificial Intelligence}} and {{Thirty-Fifth Conference}} on {{Innovative
  Applications}} of {{Artificial Intelligence}} and {{Thirteenth Symposium}} on
  {{Educational Advances}} in {{Artificial Intelligence}}}, volume~37 of
  \emph{{{AAAI}}'23/{{IAAI}}'23/{{EAAI}}'23}, pp.\  15179--15187. AAAI Press.
\newblock ISBN 978-1-57735-880-0.
\newblock \doi{10.1609/aaai.v37i12.26771}.
\newblock URL \url{https://doi.org/10.1609/aaai.v37i12.26771}.

\bibitem[Szyller et~al.()Szyller, Atli, Marchal, and
  Asokan]{szyllerDAWNDynamicAdversarial2021}
Szyller, S., Atli, B.~G., Marchal, S., and Asokan, N.
\newblock {{DAWN}}: {{Dynamic Adversarial Watermarking}} of {{Neural
  Networks}}.
\newblock In \emph{Proceedings of the 29th {{ACM International Conference}} on
  {{Multimedia}}}, {{MM}} '21, pp.\  4417--4425. Association for Computing
  Machinery.
\newblock ISBN 978-1-4503-8651-7.
\newblock \doi{10.1145/3474085.3475591}.
\newblock URL \url{https://dl.acm.org/doi/10.1145/3474085.3475591}.

\bibitem[Touvron et~al.({\natexlab{a}})Touvron, Lavril, Izacard, Martinet,
  Lachaux, Lacroix, Rozière, Goyal, Hambro, Azhar, Rodriguez, Joulin, Grave,
  and Lample]{touvronLLaMAOpenEfficient2023}
Touvron, H., Lavril, T., Izacard, G., Martinet, X., Lachaux, M.-A., Lacroix,
  T., Rozière, B., Goyal, N., Hambro, E., Azhar, F., Rodriguez, A., Joulin,
  A., Grave, E., and Lample, G.
\newblock {{LLaMA}}: {{Open}} and {{Efficient Foundation Language Models}},
  {\natexlab{a}}.
\newblock URL \url{http://arxiv.org/abs/2302.13971}.

\bibitem[Touvron et~al.({\natexlab{b}})Touvron, Martin, Stone, Albert,
  Almahairi, Babaei, Bashlykov, Batra, Bhargava, Bhosale, Bikel, Blecher,
  Ferrer, Chen, Cucurull, Esiobu, Fernandes, Fu, Fu, Fuller, Gao, Goswami,
  Goyal, Hartshorn, Hosseini, Hou, Inan, Kardas, Kerkez, Khabsa, Kloumann,
  Korenev, Koura, Lachaux, Lavril, Lee, Liskovich, Lu, Mao, Martinet, Mihaylov,
  Mishra, Molybog, Nie, Poulton, Reizenstein, Rungta, Saladi, Schelten, Silva,
  Smith, Subramanian, Tan, Tang, Taylor, Williams, Kuan, Xu, Yan, Zarov, Zhang,
  Fan, Kambadur, Narang, Rodriguez, Stojnic, Edunov, and
  Scialom]{touvronLlamaOpenFoundation2023}
Touvron, H., Martin, L., Stone, K., Albert, P., Almahairi, A., Babaei, Y.,
  Bashlykov, N., Batra, S., Bhargava, P., Bhosale, S., Bikel, D., Blecher, L.,
  Ferrer, C.~C., Chen, M., Cucurull, G., Esiobu, D., Fernandes, J., Fu, J., Fu,
  W., Fuller, B., Gao, C., Goswami, V., Goyal, N., Hartshorn, A., Hosseini, S.,
  Hou, R., Inan, H., Kardas, M., Kerkez, V., Khabsa, M., Kloumann, I., Korenev,
  A., Koura, P.~S., Lachaux, M.-A., Lavril, T., Lee, J., Liskovich, D., Lu, Y.,
  Mao, Y., Martinet, X., Mihaylov, T., Mishra, P., Molybog, I., Nie, Y.,
  Poulton, A., Reizenstein, J., Rungta, R., Saladi, K., Schelten, A., Silva,
  R., Smith, E.~M., Subramanian, R., Tan, X.~E., Tang, B., Taylor, R.,
  Williams, A., Kuan, J.~X., Xu, P., Yan, Z., Zarov, I., Zhang, Y., Fan, A.,
  Kambadur, M., Narang, S., Rodriguez, A., Stojnic, R., Edunov, S., and
  Scialom, T.
\newblock Llama 2: {{Open Foundation}} and {{Fine-Tuned Chat Models}},
  {\natexlab{b}}.
\newblock URL \url{http://arxiv.org/abs/2307.09288}.

\bibitem[Tramèr \& Boneh()Tramèr and Boneh]{tramerSlalomFastVerifiable2019}
Tramèr, F. and Boneh, D.
\newblock Slalom: {{Fast}}, {{Verifiable}} and {{Private Execution}} of
  {{Neural Networks}} in {{Trusted Hardware}}.
\newblock URL \url{http://arxiv.org/abs/1806.03287}.

\bibitem[Tramèr et~al.()Tramèr, Zhang, Juels, Reiter, and
  Ristenpart]{tramerStealingMachineLearning2016}
Tramèr, F., Zhang, F., Juels, A., Reiter, M.~K., and Ristenpart, T.
\newblock Stealing {{Machine Learning Models}} via {{Prediction}}
  \{\vphantom\}{{APIs}}\vphantom\{\}.
\newblock pp.\  601--618.
\newblock ISBN 978-1-931971-32-4.
\newblock URL
  \url{https://www.usenix.org/conference/usenixsecurity16/technical-sessions/presentation/tramer?}

\bibitem[Tsai et~al.()Tsai, Porter, and
  Vij]{tsaiGrapheneSGXPracticalLibrary2017}
Tsai, C.-C., Porter, D.~E., and Vij, M.
\newblock \{\vphantom\}{{Graphene-SGX}}\vphantom\{\}: {{A Practical Library}}
  \{\vphantom\}{{OS}}\vphantom\{\} for {{Unmodified Applications}} on
  \{\vphantom\}{{SGX}}\vphantom\{\}.
\newblock pp.\  645--658.
\newblock ISBN 978-1-931971-38-6.
\newblock URL
  \url{https://www.usenix.org/conference/atc17/technical-sessions/presentation/tsai}.

\bibitem[Viand \& Shafagh()Viand and Shafagh]{viandMarbleMakingFully2018a}
Viand, A. and Shafagh, H.
\newblock Marble: {{Making Fully Homomorphic Encryption Accessible}} to
  {{All}}.
\newblock In \emph{Proceedings of the 6th {{Workshop}} on {{Encrypted
  Computing}} \& {{Applied Homomorphic Cryptography}}}, {{WAHC}} '18, pp.\
  49--60. Association for Computing Machinery.
\newblock ISBN 978-1-4503-5987-0.
\newblock \doi{10.1145/3267973.3267978}.
\newblock URL \url{https://doi.org/10.1145/3267973.3267978}.

\bibitem[Voss()]{vossEuropeanUnionData2017}
Voss, W.~G.
\newblock European {{Union Data Privacy Law Reform}}: {{General Data Protection
  Regulation}}, {{Privacy Shield}}, and the {{Right}} to {{Delisting}}.
\newblock URL \url{https://papers.ssrn.com/abstract=2894571}.

\bibitem[Wood et~al.()Wood, Najarian, and
  Kahrobaei]{woodHomomorphicEncryptionMachine2020}
Wood, A., Najarian, K., and Kahrobaei, D.
\newblock Homomorphic {{Encryption}} for {{Machine Learning}} in {{Medicine}}
  and {{Bioinformatics}}.
\newblock 53\penalty0 (4):\penalty0 70:1--70:35.
\newblock ISSN 0360-0300.
\newblock \doi{10.1145/3394658}.
\newblock URL \url{https://dl.acm.org/doi/10.1145/3394658}.

\bibitem[Wu et~al.({\natexlab{a}})Wu, Fei, Qu, Ji, and
  Chua]{wuNExTGPTAnytoAnyMultimodal2023}
Wu, S., Fei, H., Qu, L., Ji, W., and Chua, T.-S.
\newblock {{NExT-GPT}}: {{Any-to-Any Multimodal LLM}}, {\natexlab{a}}.
\newblock URL \url{http://arxiv.org/abs/2309.05519}.

\bibitem[Wu et~al.({\natexlab{b}})Wu, Irsoy, Lu, Dabravolski, Dredze, Gehrmann,
  Kambadur, Rosenberg, and Mann]{wuBloombergGPTLargeLanguage2023}
Wu, S., Irsoy, O., Lu, S., Dabravolski, V., Dredze, M., Gehrmann, S., Kambadur,
  P., Rosenberg, D., and Mann, G.
\newblock {{BloombergGPT}}: {{A Large Language Model}} for {{Finance}},
  {\natexlab{b}}.
\newblock URL \url{http://arxiv.org/abs/2303.17564}.

\bibitem[Xue et~al.({\natexlab{a}})Xue, Wu, He, Wang, and
  Liu]{xueActiveDNNIP2020}
Xue, M., Wu, Z., He, C., Wang, J., and Liu, W.
\newblock Active {{DNN IP Protection}}: {{A Novel User Fingerprint Management}}
  and {{DNN Authorization Control Technique}}.
\newblock In \emph{2020 {{IEEE}} 19th {{International Conference}} on
  {{Trust}}, {{Security}} and {{Privacy}} in {{Computing}} and
  {{Communications}} ({{TrustCom}})}, pp.\  975--982, {\natexlab{a}}.
\newblock \doi{10.1109/TrustCom50675.2020.00130}.
\newblock URL \url{https://ieeexplore.ieee.org/abstract/document/9343023}.

\bibitem[Xue et~al.({\natexlab{b}})Xue, Zhang, Wang, and
  Liu]{xueIntellectualPropertyProtection2022}
Xue, M., Zhang, Y., Wang, J., and Liu, W.
\newblock Intellectual {{Property Protection}} for {{Deep Learning Models}}:
  {{Taxonomy}}, {{Methods}}, {{Attacks}}, and {{Evaluations}}.
\newblock 3\penalty0 (06):\penalty0 908--923, {\natexlab{b}}.
\newblock ISSN 2691-4581.
\newblock \doi{10.1109/TAI.2021.3133824}.
\newblock URL
  \url{https://www.computer.org/csdl/journal/ai/2022/06/09645219/1zc6IpBDlfi}.

\bibitem[Zhao et~al.()Zhao, Zhou, Li, Tang, Wang, Hou, Min, Zhang, Zhang, Dong,
  Du, Yang, Chen, Chen, Jiang, Ren, Li, Tang, Liu, Liu, Nie, and
  Wen]{zhaoSurveyLargeLanguage2023}
Zhao, W.~X., Zhou, K., Li, J., Tang, T., Wang, X., Hou, Y., Min, Y., Zhang, B.,
  Zhang, J., Dong, Z., Du, Y., Yang, C., Chen, Y., Chen, Z., Jiang, J., Ren,
  R., Li, Y., Tang, X., Liu, Z., Liu, P., Nie, J.-Y., and Wen, J.-R.
\newblock A {{Survey}} of {{Large Language Models}}.
\newblock URL \url{http://arxiv.org/abs/2303.18223}.

\end{thebibliography}
\bibliographystyle{icml2024}

\end{document}